\documentclass[twocolumn,tighten]{aastex62}
\usepackage{bm}
\usepackage{here}
\received{\today}
\revised{ }
\accepted{ }
\submitjournal{ApJ}

\shorttitle{Internal Rotation of KIC11145123}
\shortauthors{Hatta et al.}

\begin{document}

\title{The two-dimensional internal rotation of KIC11145123\footnote{Released on January, Xth, 20XX}}

\correspondingauthor{Yoshiki Hatta}
\email{yoshiki.hatta@nao.ac.jp}


\author{Yoshiki Hatta}
\affiliation{Department of Astronomical Science, School of Physical Sciences, SOKENDAI\\
2-21-1 Osawa, Mitaka, Tokyo 181-8588, Japan}
\affiliation{National Astronomical Observatory of Japan \\
2-21-1 Osawa, Mitaka, Tokyo 181-8588, Japan}

\author{Takashi Sekii}
\affiliation{Department of Astronomical Science, School of Physical Sciences, SOKENDAI\\
2-21-1 Osawa, Mitaka, Tokyo 181-8588, Japan}
\affiliation{National Astronomical Observatory of Japan \\
2-21-1 Osawa, Mitaka, Tokyo 181-8588, Japan}


\author{Masao Takata}
\affiliation{Department of Astronomy, School of Science, The University of Tokyo\\
Bunkyou-ku, Tokyo 113-0033, Japan}

\author{Donald W. Kurtz}
\affiliation{Jeremiah Horrocks Institute, University of Central Lancashire\\
Preston PR1 2HE, UK}


%
%
%



\begin{abstract}
The two-dimensional internal rotation of KIC11145123 has been inferred via asteroseismology. Based on the Optimally Localized Averaging method and a simple three-zone modeling of the internal rotation, we have found evidence for a contrast between the internal rotation of the radiative region and that of the convective core; the radiative region rotates almost uniformly throughout the region, but the convective core may be rotating about 6 times faster than the radiative region above. We have also found a marginally significant evidence of latitudinal differential rotation in the outer envelope. These newly indicated features of the internal rotation of the star can help us further constrain the theory of angular momentum transport inside stars as well as understand the complex physical properties of the star, which was once thought to be a main-sequence A-type star but recently has been proposed to be a blue straggler, based on spectroscopy.

%
\end{abstract}



\section{Introduction} \label{sec:intro}
Stellar rotation has been an important topic in the field of stellar physics for a long time. Specifically, stellar internal rotation has considerable influences on a variety of physical processes inside stars \citep[e.g.][]{Maeder2009}. For instance, it is generally considered that angular velocity shear induces an extra mixing inside the star, modifies the distributions of chemical elements, and thus, has a strong impact on the stellar evolution. Another example is generation of magnetic fields inside stars. Stellar differential rotation can directly convert a poloidal magnetic field into a toroidal magnetic field ($\Omega$-effect), contributing to dynamo processes inside the star. Hence, observationally inferring stellar internal rotation is a critical key in order for us to disentangle complicated physical processes in stars which occur with a broad range of timescales.

We are able to estimate the internal rotation of stars via asteroseismology \citep[e.g.][]{Aerts2010}, which is a branch of stellar physics where we infer interiors of stars based on the measurements of stellar oscillations. In particular, if a star is rotating so slowly that we can neglect the second-order effects of rotation on the eigenfrequencies, frequency shifts caused by the rotation, known as rotational splitting, are proportional to internal rotation rates averaged over regions where the corresponding eigenmodes have sensitivity \citep{Ledoux1951}. The first asteroseismic inference of stellar internal rotation based on the first-order perturbative method was carried out for white dwarfs \citep{Kawaler1999} and subsequently for main-sequence B stars \citep{Aerts2003, Briquet2007}. However, the former study only suggested the existence of the differential rotation, and the latter only derived ratios of the angular velocity in the core to that in the envelope; the small number of detected modes made it difficult to draw more spatially resolved information on the internal rotation of the stars in both cases. There are also uncertainties in their mode identification, leading to a difficulty in eliminating the model dependence of their inferences. These situations have been changed with the advent of the space-borne missions such as CoRoT \citep{CoRoT} and Kepler \citep{Kepler}. Nowadays, because of the unprecedented precision of the frequency measurements, there have been numerous studies that have revealed the internal rotation of stars including main-sequence B, A, F, and G type stars \citep[e.g.][]{Kurtz2014, Saio2015, Benomar2015, Schmid2016, Van Reeth2016, Papics2017, Christophe2018, Ouazzani2018} and red-giant stars \citep[e.g.][]{Beck2012, Deheuvels2012, Mosser2012}. 

\citet{Aerts2017} summarized the current understanding of the internal rotation of main-sequence A-F stars to show that almost all the stars that have been investigated so far are nearly rigidly rotating regardless of their rotation periods, suggesting that there should be strong mechanisms of angular momentum transport throughout the radiative regions of the stars. However, these results are not consistent with the predictions of the current theory of angular momentum transport just with hydrodynamical processes \citep[e.g.][]{Tayar2013, Eggenberger2017}, indicating that we have to take into account additional mechanisms such as the transport by magnetic fields \citep{Cantiello2014} and internal gravity waves \citep{Rogers2015}. Though there have been no unanimous conclusions yet, it is obvious that asteroseismology has successfully put constraints on the theory of angular momentum transport.

Among the most interesting stars that have been asteroseismically analyzed until now is KIC11145123, which was thought to be a main-sequence A-type star, with the surprisingly slow rotation period for a typical A star, $P_{\rm{rot}}\sim100 \ \rm{d}$ \citep{Kurtz2014}. The star exhibits pressure modes (p modes) and gravity modes (g modes) \citep{Kurtz2014}, presenting the possibility that we can infer both the outer envelope and the inner core; as p modes probe outer regions and g modes probe inner regions in the case of an ordinary main-sequence star \citep[e.g.][section 14]{Unno1989}. Furthermore, both p- and g-mode rotational splitting are well resolved, and thus, we can infer the internal rotation of the inner region and that of the outer region at the same time with high precision. There are only two other such stars \citep{Saio2015, Schmid2016}. Although there have been a number of hybrid stars found by Kepler \citep{Bradley2015}, the star is also exceptional in terms of its unique internal rotation profile. \citet{Kurtz2014} inferred the one-dimensional (latitudinally averaged) internal rotation based on the model-independent consideration of the observed rotational splitting, and they revealed that the core of the star rotates slightly slower than the envelope does. During the evolution on the main sequence, a star's core gradually shrinks and the envelope expands; the core of the star should rotate much faster than the envelope if the angular momentum is locally conserved, but this is not the case for KIC11145123. \citet{Kurtz2014} suggested that the star is a blue straggler \citep{Sandage1953} and it has obtained an extra angular momentum from the outside which has led to the spin-up of the outer envelope. Interestingly, \citet{Takada-Hidai2017} have conducted spectroscopic analyses of the star with Subaru/HDS, and they have concluded that the star is spectroscopically a blue straggler, supporting the suggestion by \citet{Kurtz2014}. In addition, \citet{Gizon2016} have carried out a further detailed asteroseismic analysis to measure the asphericity of the star. What they have revealed is that the star is less oblate than expected from its rotation period. 

Still, the origin of the rotation profile remains uncertain, and there is a possibility that the different trend of the internal rotation can be found, for instance, in high-latitude region of the star; \citet{Kurtz2014} inferred the internal rotation of the star assuming that the internal rotation is not dependent on latitude. Thus, there is interest in carrying out rotation inversion taking latitudinal dependence into account, from which we hope to extract more spatially resolved information on the internal rotation. It is also essential to grasp such detailed rotation profile since we could further put a constraint on the theory of stellar rotation. Furthermore, this is the first attempt to investigate the two-dimensional internal rotation of a star other than the Sun. \citet{Lund2014} suggested the possibility that we can asteroseismically infer the latitudinal differential rotation for solar-like stars, and \citet{Benomar2018} have recently reported the first detection of the latitudinal differential rotation for 13 solar-like stars observed by Kepler. However, two-dimensional rotation profiles of the stars have not been studied yet.

The goal of this paper is to infer the two-dimensional internal rotation profile of KIC11145123 by conducting rotation inversion. The structure of this paper is as follows. In Section \ref{2}, we take a look at the basic settings of the equilibrium model of the star, the frequency analysis, and the calculation of linear adiabatic oscillation. Then, in the next section, we show the principle of the first-order perturbative method which relates rotational splitting to the internal rotation, and we also present how to estimate internal rotation based on the set of linear integral constraints. We describe the results of the inversion and discuss them in Sections \ref{4} and \ref{5}, respectively. Finally, we present a summary in Section \ref{6}.

\section{Observation and Model}
\label{2}
The data and the equilibrium model that we have used are based on those in \citet{Kurtz2014}. We present the basic settings below.

\subsection{Frequency analysis}
\label{2-1}
\citet{Kurtz2014} firstly measured the frequencies of the 76 detected modes (45 g modes and 31 p modes) \citep[see Figure 1 of][]{Kurtz2014} by a non-linear least-squares method. The estimated errors are as small as the order of $10^{-6}$ cycle per day due to the coherence of the pulsations. It is evident that the rotational splitting occurs \citep[see Figure 3 of][]{Kurtz2014}, and the values of the rotational shifts in frequencies, or ``generalized rotational splittings'' \citep{Goupil2011}, are simply computed as below:
\begin{equation}
\Delta \omega_{nlm} = \frac{\omega_{nl,m}-\omega_{nl,-m}}{2}, \label{equation0}
\end{equation}
where $n$, $l$, and $m$ are the radial order, the spherical degree, and the azimuthal order, respectively. We have used the rotational shifts and the corresponding errors estimated by \citet{Kurtz2014} in the following analyses.

We also see approximately constant g-mode period spacing which is expected from asymptotic behavior of high-order g modes \citep{Tassoul1980}, and the observed value is $\Delta P_{\rm{g}} = 0.0245 \  \rm{d}$. \citet{Kurtz2014} fitted the observed mean g-mode period spacing to produce their best model of the star (see Section \ref{2-2}). For more information on the individual frequencies, amplitudes, phases, and estimated errors, see \citet{Kurtz2014}.

\subsection{Equilibrium model and linear adiabatic oscillation}
\label{2-2}
\begin{figure} [t]
\begin{center}
 \includegraphics[scale=0.52]{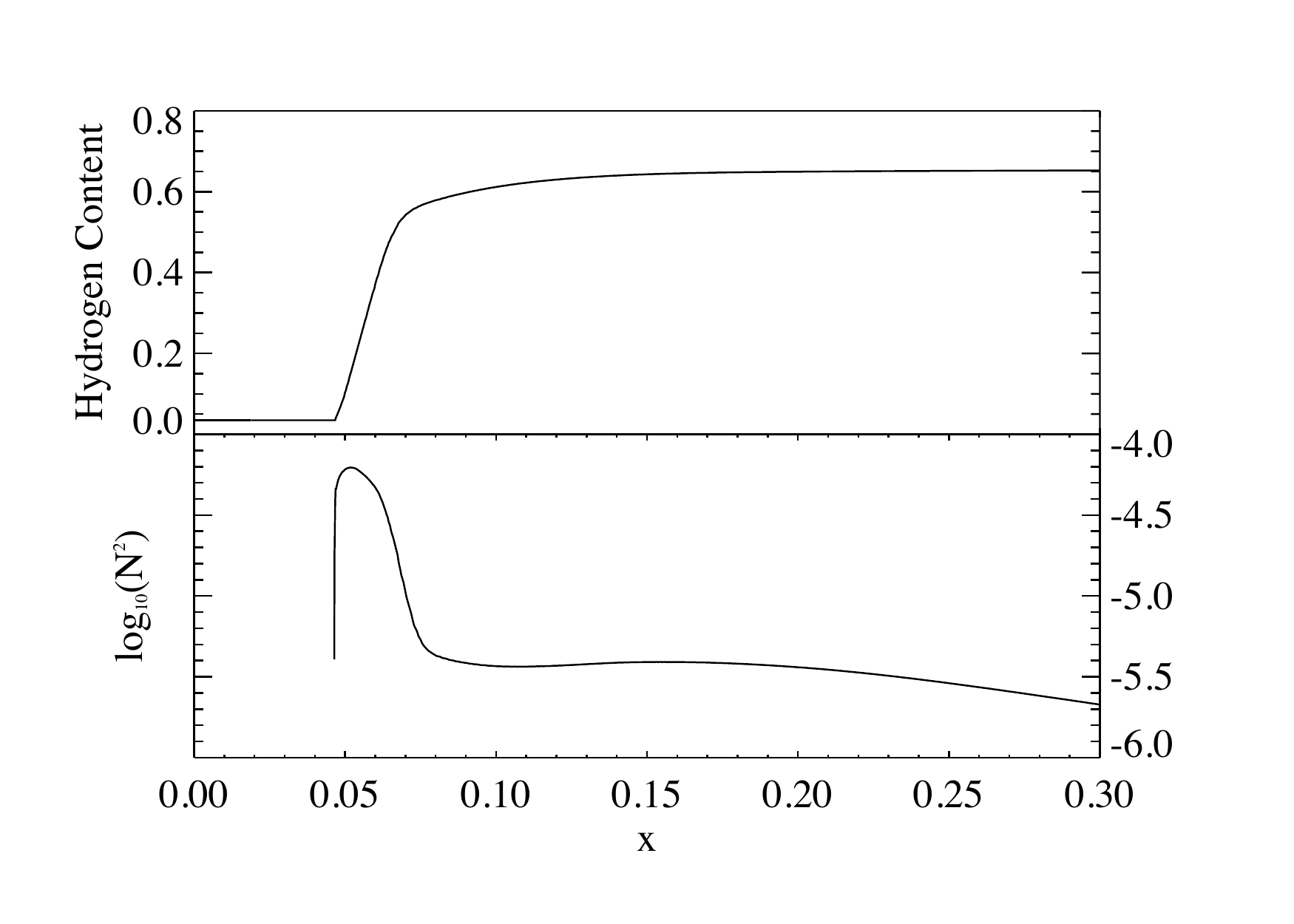}
 \caption{\footnotesize The hydrogen mass fraction (top) and the logarithm of the squared Brunt-$\rm{V}\ddot{a}is\ddot{a}l\ddot{a}$ frequency in units of Hz (bottom) of the reference model in the present study. It is obvious that the star is at TAMS stage and it has nearly exhausted hydrogen in the core region ($x<0.045$). The square of the Brunt-$\rm{V}\ddot{a}is\ddot{a}l\ddot{a}$ frequency is negative inside the convective core where buoyancy does not act as a restoring force. A minor note is that this figure is based on a model recalculated by us via MESA, version 9793 \citep{Paxton2015}. The results are essentially unchanged.}
 \label{1.46M_internal_structures}
\end{center} 
\end{figure}

We have chosen Kurtz et al.'s model as a reference model. The model was calculated using Modules for Experiments in Stellar Astrophysics (MESA, version 4298) \citep{Paxton2013}. The initial hydrogen content $X_{\rm{init}}$, the initial helium content $Y_{\rm{init}}$, and the initial metallicity $Z_{\rm{init}}$ are 0.65, 0.34, and 0.01, respectively. The initial mass is $1.46\, M_{\odot}$. The model is found to be at the terminal age main-sequence (TAMS) stage where the star has exhausted almost all the hydrogen in the core (Figure \ref{1.46M_internal_structures}). There is a sharp feature in the Brunt-$\rm{V}\ddot{a}is\ddot{a}l\ddot{a}$ frequency and the location of the peak corresponds to the chemical composition gradient left by the receding convective core during the evolution (Figure \ref{1.46M_internal_structures}). For more details, see section 3 in \citet{Kurtz2014}. 
\begin{table} [t]
 \begin{center}
  \caption{\footnotesize The observed $m=0$ mode frequencies $f_{\rm{obs}}$ of KIC11145123 \citep{Kurtz2014}. The first column is the mode type. The radial order and the spherical degree are the third column. The observational error is the fourth column. Here, we also define the mode numbers for associated rotational shifts for discussions in Sections \ref{4} and \ref{5}.}
    \begin{tabular} {ccccc} \hline
  & $f_{\rm{obs}}$ & ($n$,$l$) & Error & Mode Number\\ 
 & ($\textrm{d}^{-1}$)  & &  ($\textrm{d}^{-1}$) & \\ \hline
  g & 1.209 5758 & (-33,1) & 0.000 0030 & 1 (m=1)\\
  g & 1.247 1853 & (-32,1) & 0.000 0127 & 2 (m=1)\\
  g & 1.287 0610 & (-31,1) & 0.000 0086 & 3 (m=1)\\
  g & 1.327 6030 & (-30,1) & 0.000 0042 & 4 (m=1)\\
  g & 1.373 5146 & (-29,1) & 0.000 0018 & 5 (m=1)\\
  g & 1.418 1646 & (-28,1) & 0.000 0015 & 6 (m=1)\\
  g & 1.469 6740 & (-27,1) & 0.000 0021 & 7 (m=1)\\
  g & 1.523 5985 & (-26,1) & 0.000 0033 & 8 (m=1)\\
  g & 1.581 5714 & (-25,1) & 0.000 0171 & 9 (m=1)\\
  g & 1.647 0217 & (-24,1) & 0.000 0068 & 10 (m=1)\\
  g & 1.712 9908 & (-23,1) & 0.000 0255 & 11 (m=1)\\
  g & 1.789 2207 & (-22,1) & 0.000 0297 & 12 (m=1)\\
  g & 1.864 0302 & (-21,1) & 0.000 0449 & 13 (m=1)\\
  g & 1.950 8535 & (-20,1) & 0.000 0302 & 14 (m=1)\\
  g & 2.044 7027 & (-19,1) & 0.000 0578 & 15 (m=1)\\
  p & 17.963 5133 & (3,0) & 0.000 0005 & \\
  p & 18.366 0001 & (2,1) & 0.000 0135 & 16 (m=1)\\
  p & 22.001 8915 & (3,1) & 0.000 0286 & 17 (m=1)\\
  mixed & 16.742 0104 & (-1,2) & 0.000 0075 & 18 (m=1)\\
             &                     &                                 &                     & 21 (m=2)\\
  mixed & 19.006 4482 & (0,2) & 0.000 0056 & 19 (m=1)\\
             &                     &                                 &                     & 22 (m=2)\\
  mixed & 23.565 1835 & (2,2) & 0.000 0480 & 20 (m=1)\\
               &                     &                                 &                     & 23 (m=2)\\  \hline

   \end{tabular}
 \label{freqs_modeled}
\end{center}
\end{table}

\citet{Kurtz2014} also calculated linear adiabatic oscillation of the model based on the pulsation codes developed by \citet{Saio1980} and \citet{Takata2012}, and they conducted mode identification of the detected modes by comparing the modeled frequencies with the observed frequencies (Table \ref{freqs_modeled}). In our analyses, we rely on their identification and we have used the eigenfunctions they computed, in calculating 2-dimensional rotational splitting kernel that relates the rotational shifts to internal rotation (see Section \ref{3}).

As a final remark on the model, we would like to point out that neither several p-mode frequencies nor the atmospheric parameters such as $T_{\rm{eff}}$ and $\rm{log} \ \it{g}$ are reproduced well with the current model. This could be partly because the star is probably a blue straggler, having experienced some interactions with other stars during its evolution; modeling the star as an ordinary A star may not be sufficient. In fact, \citet{Takada-Hidai2017} have carried out spectroscopic analyses of the star with Subaru/HDS and they found that it is spectroscopically a blue straggler. We will tackle the problem based on non-standard modeling of the star in our forthcoming paper and we assume that the current model is good enough to explain KIC11145123 in the rest of this paper. Actually, we have confirmed that the results of rotation estimation are not dramatically changed even though we used other models such as those produced by \citet{Takada-Hidai2017} and by Hatta et al., in prep.

\section{Method}
\label{3}
The internal rotation of a star can be inferred from the first-order perturbative method \citep[e.g.][]{Thompson1996}, if the star is rotating so slowly that we can neglect the second-order effects on the oscillation \citep{Ledoux1951}. This is the case for KIC11145123 with a rotation period, $P_{\rm{rot}}\sim 100 \  \rm{d}$, which is much longer than the dynamical timescale of the star, $\tau_{\rm{dyn}}\sim 0.06\ \rm{d}$, as well as the oscillation periods of the star, $P_{\rm{osc}}\sim 0.05$-$1 \  \rm{d}$. Firstly, we would like to describe how the effects of the rotation on the eigenfrequencies are expressed to the first-order, and we also show the splitting kernels which relate the internal rotation to the rotational shifts. Then, we present how to estimate the internal rotation using the relation.

\subsection{Rotational splitting}
\label{3-1}
\begin{figure} [t]
\begin{center}
\includegraphics[scale=0.5]{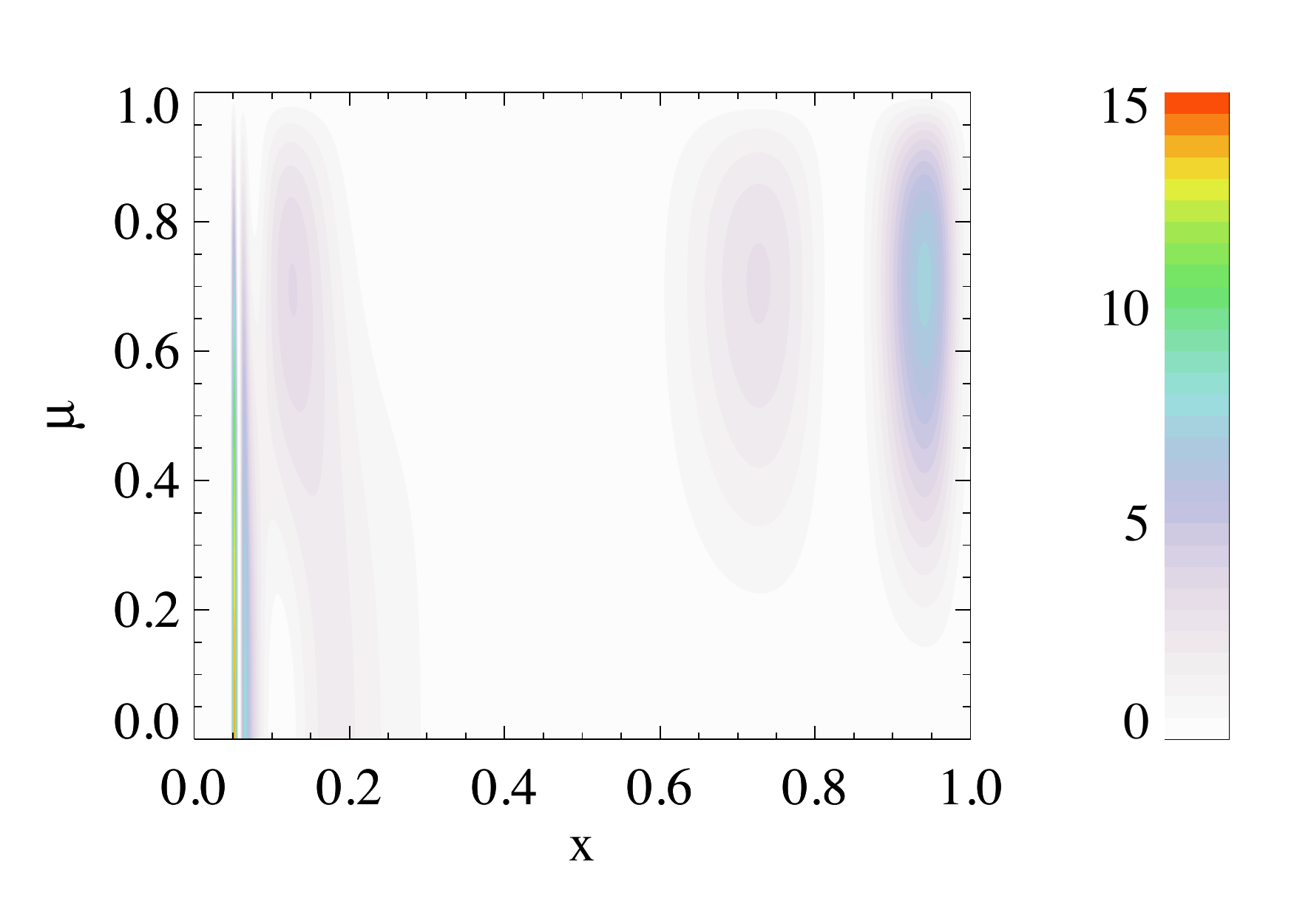}
\caption{\footnotesize Splitting kernel of the mixed mode with $(n,l,m)=(-1,2,1)$. In the outer envelope, the kernel has maxima at middle latitude (around $\mu = 0.7$), suggesting the possibility that we can extract the information on the internal rotation at high latitudes. In the central region, on the other hand, the kernel is centered around the equatorial plane because of the dominancy of the horizontal component of the eigenfunction over the radial component \citep[see more details in][]{Sekii1995}.}
\label{rotation_kernel_p_nminus1_l2_m1.png}
\end{center} 
\end{figure}

\begin{figure} [t]
\begin{center}
\includegraphics[scale=0.5]{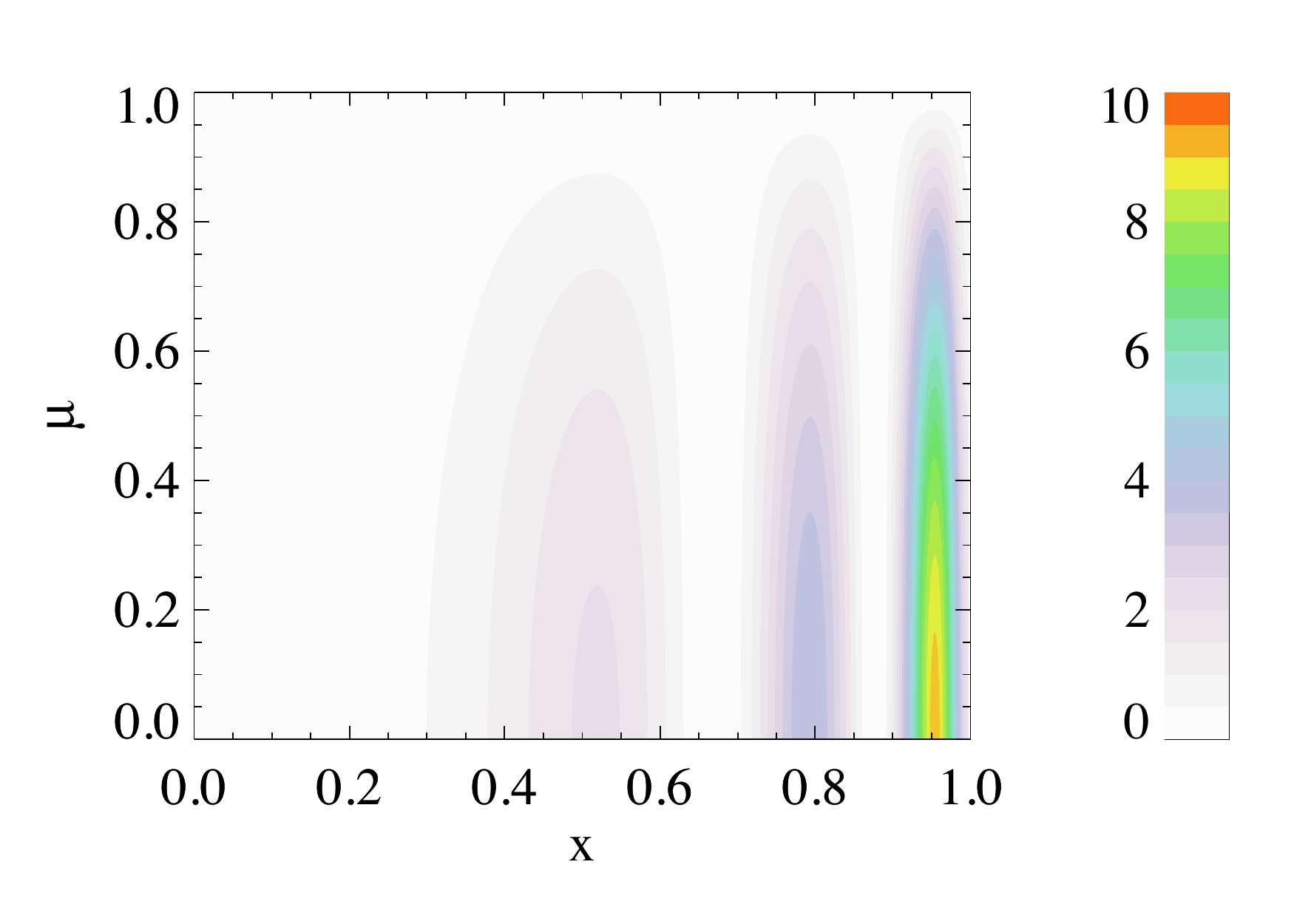}
\caption{\footnotesize Splitting kernel of the p mode with $(n,l,m)=(2,1,1)$. It actually has a very slight g-mode nature, but the sensitivity of the kernel is mainly concentrated in the outer envelope.}
\label{rotation_kernel_p_n2_l1.png}
\end{center} 
\end{figure}

\begin{figure} [t]
\begin{center}
\includegraphics[scale=0.5]{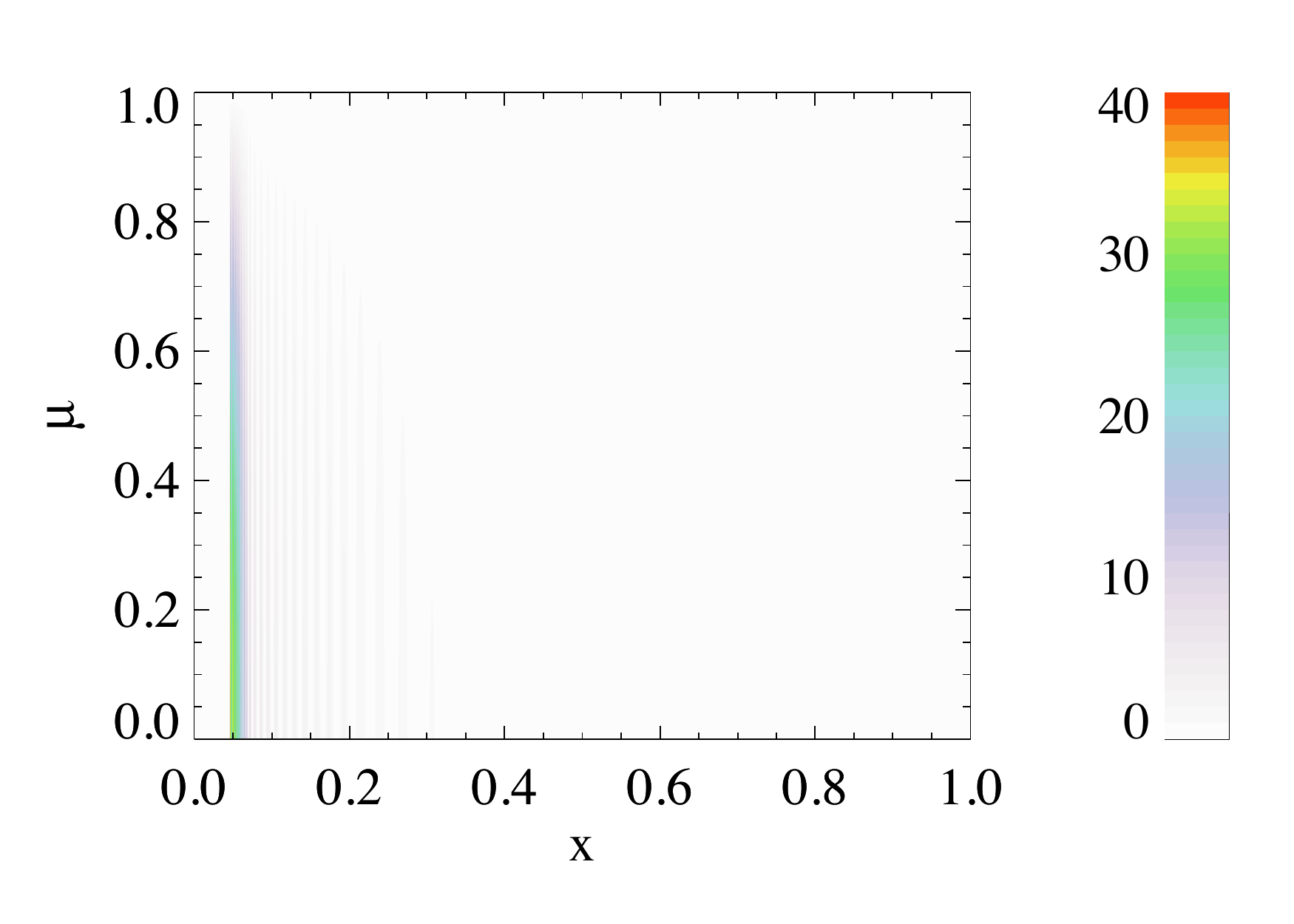}
\caption{\footnotesize Splitting kernel of the g mode with $(n,l,m)=(-33,1,1)$. The sensitivity is concentrated on just above the edge of the convective core around $x=0.045$. There is also a latitudinal dependence which is determined by associated Legendre function \citep[see more details in][]{Sekii1995}.}
\label{rotation_kernel_g_nminus33_l1.png}
\end{center} 
\end{figure}

When the internal rotation is slow enough and we can treat it as a small perturbation, perturbation theory tells us that the rotational shift, $\Delta \omega_{i}$ with $i$ representing the mode indices $(n,l,m)$, is related to the internal rotation profile $\Omega(x,\mu)$ by the splitting kernel $K_{i}(x,\mu)$ as follows:
\begin{equation}
d_{i} = \int \!\!\! \int K_{i}(x,\mu) \Omega (x,\mu) dx d \mu + e_{i}, \; \; \; \; i=1,...,M, \label{equation5}
\end{equation}
where $x$ and $\mu$ are fractional radius and cosine of the colatitude, respectively. We have replaced $\Delta \omega_{i}/m$ with $d_i$ and $M$ is the number of the observed rotational shifts used in the rotation inversion, namely, 23 in the case of this study. Observational errors $e_{i}$ are included in expression (\ref{equation5}). Relation (\ref{equation5}) is correct to first-order. The splitting kernels can be calculated once we have solved the linear adiabatic oscillation of the equilibrium model \citep[e.g.][]{Sekii1995}. Note that relation (\ref{equation5}) shows that the rotational shifts can be interpreted as averaged values of the internal angular velocity weighted by the splitting kernel. In other words, the splitting kernels give us information about regions where the eigenfrequencies are most influenced by the internal rotation there (Figures \ref{rotation_kernel_p_nminus1_l2_m1.png} to \ref{rotation_kernel_g_nminus33_l1.png}). 

We have this form of equation for each observed rotational shift, and thus we can estimate the internal rotation $\Omega(x,\mu)$.

\subsection{Optimally Localized Averaging method}
\label{3-2}
What we have to do is to estimate the internal rotation profile $\Omega(x,\mu)$ based on the set of linear integral constraints (\ref{equation5}).

The basic point of the Optimally Localized Averaging (OLA) method is that we express an estimate at a certain target point $(x_{0},\mu_{0})$ as a linear combination of data $d_{i}$
\begin{eqnarray}
& \hat{ \Omega } &(x_{0},\mu_{0}) = \sum_{i=1}^{M} c_{i}(x_{0},\mu_{0}) d_{i} \nonumber  \\
  & = & \int \!\!\! \int D(x,\mu ; x_{0}, \mu_{0}) \Omega (x, \mu) dxd\mu + \sum_{i=1}^{M} c_{i}(x_{0},\mu_{0}) e_{i}, \nonumber \\
   \label{equation6}
\end{eqnarray}
where $D(x,\mu ; x_{0}, \mu_{0})$ is the averaging kernel, defined as 
\begin{equation}
D(x,\mu ; x_{0}, \mu_{0}) \equiv \sum_{i=1}^{M} c_{i}(x_{0}, \mu_{0})K_{i}(x,\mu) . \label{equation7}
\end{equation}

We can obtain a good estimate of the internal rotation at the target point by making the averaging kernel as localized as possible around the target point. For the purpose of localization, it is common that we minimize the following quantity
\begin{equation}
S =  \int \!\!\! \int D(x,\mu ; x_{0}, \mu_{0})^2 \biggl \{ (x-x_{0})^2 + x^2(\mu - \mu_{0})^2 \biggr \} dx d\mu,  \label{equation9}
\end{equation}
with the unimodular condition of $D(x,\mu;x_{0}, \mu_{0})$
\begin{equation}
S_{u} \equiv  \int \!\!\! \int D(x,\mu ; x_{0}, \mu_{0})  dx d\mu = 1.  \label{equation9}
\end{equation}
Minimizing $S$ forces the averaging kernel to be large near the target point and to be small far from it \citep{BackusGilbert1967}. On the other hand, we would like to keep the estimated errors small at the same time. Thus, what we actually minimize is 
\begin{equation}
S_{\alpha} = S + \alpha \sum_{i,j=1}^{M}c_i c_j E_{ij} + 2 c_{M+1} (1-S_{u} ),  \label{equation10}
\end{equation}
where $\alpha$ is a trade-off parameter which determines the balance between the first term in expression (\ref{equation10}), resolution, and the second term, error magnification. The unimodular condition of the averaging kernel is expressed by using a Lagrangian multiplier $c_{M+1}$ introduced as the $(M+1)$-th coefficient. The error covariance matrix is denoted by $E_{ij}$. In this study, we assume that the observational error is independent of each other, in other words, $E_{ij}=e_{i}^2 \delta_{ij}$.

The minimization condition can be rewritten using matrices, and we obtain the inversion coefficients $c_{i}(x_{0},\mu_{0})$ by inverting the linear matrix equation. We finally obtain the estimate for the target point $(x_{0},\mu_{0})$ by substituting $c_{i}$ for the first line in the expression (\ref{equation6}).
\begin{figure} [hbt]
\begin{center}
\includegraphics[scale=0.5]{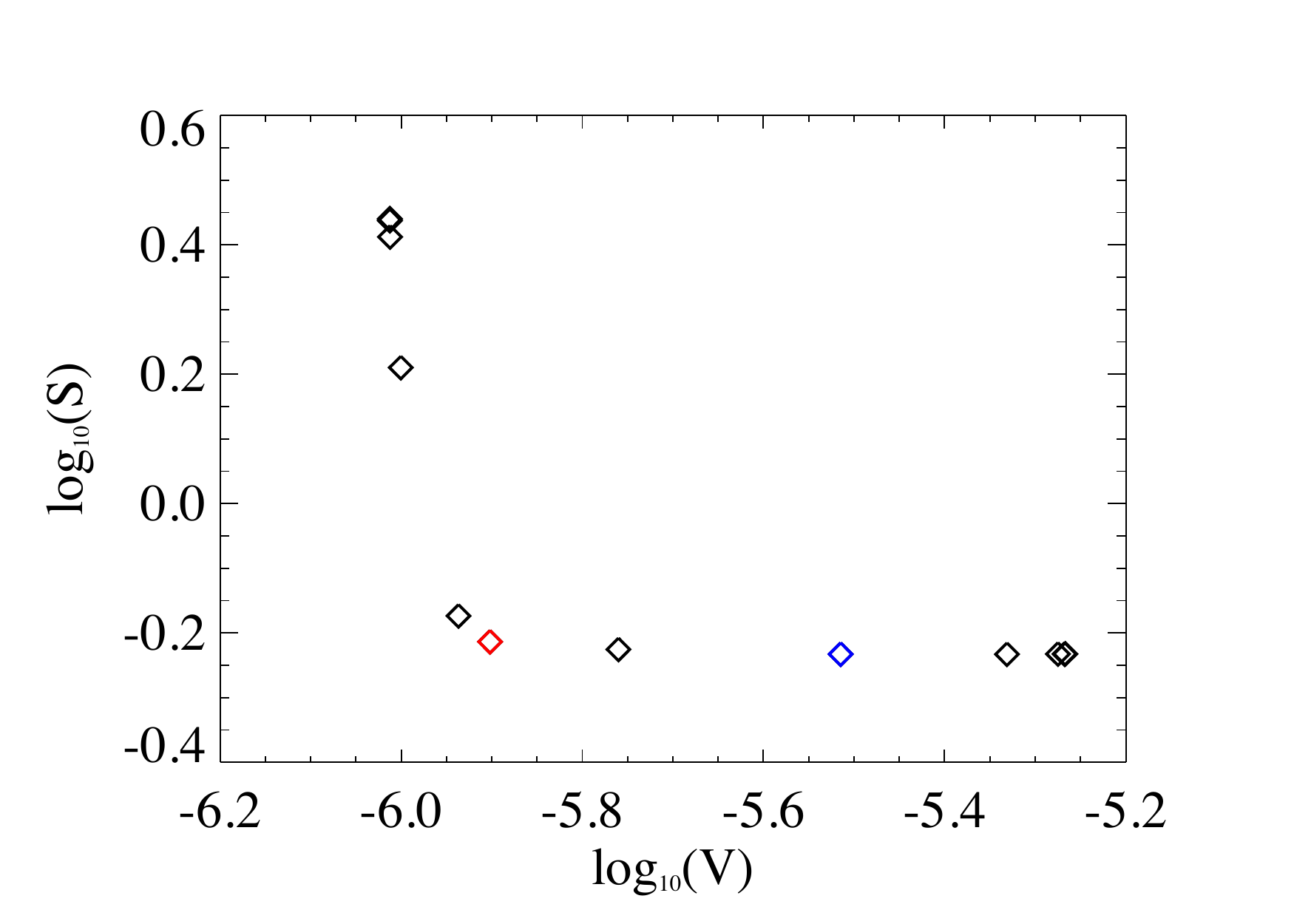}
\caption{\footnotesize Trade-off curve in the case of rotation inversion with a target point $(x,\mu)=(0.05,0.00)$. It is easily seen that error magnification $V$ becomes larger as the width of the averaging kernels $S$ becomes smaller, and vice versa. The red and blue diamonds show the values of $(S,V)$ with which the estimates (\ref{estimate3-2}) and (\ref{estimate3-1}) are obtained, respectively. } 
\label{2dOLA_trade_off_xt005_mut000.png}
\end{center} 
\end{figure}

\subsection{Three target points}
\label{3-3}
Because the averaging kernel is just a linear combination of the splitting kernels, where we can localize the averaging kernel is determined by the shapes of the splitting kernels. We present three examples of the splitting kernels (Figures \ref{rotation_kernel_p_nminus1_l2_m1.png} to \ref{rotation_kernel_g_nminus33_l1.png}), and as we see, there are three regions where the splitting kernels have good sensitivity: the outer region at middle latitude (Figure \ref{rotation_kernel_p_nminus1_l2_m1.png}), the outer region at low latitude (Figure \ref{rotation_kernel_p_n2_l1.png}), and the deep radiative region at low latitude (Figure \ref{rotation_kernel_g_nminus33_l1.png}). We have chosen the three regions as the target points, namely, $(x,\mu)=(0.05,0.00), (0.95,0.00)$, and $(0.95, 0.70)$, in the following inversion via the OLA method. Note that when we selected the target points far from these regions we were not able to localize the averaging kernels very well, leading to unreliable estimates.

\section{Result}
\label{4}
We show the results of rotation inversion via the OLA method. We have estimated the internal rotation for the three target points that have been selected in the previous section.

\subsection{Internal rotation at the three target points}
\label{4-1}
Before we show the estimates of the internal rotation, we briefly explain how we have determined the value of the trade-off parameter $\alpha$ (see equation (\ref{equation10})); we must select $\alpha$ in order to obtain estimates. For this purpose, we have repeated the rotation inversion $40$ times for different values of $\alpha$, and then, quantified the width of averaging kernels $S$, which corresponds to the first term in expression (\ref{equation10}), and error magnification $V$, which corresponds to the second term divided by alpha in the same expression (\ref{equation10}). We have plotted the trade-off curves (e.g. Figure \ref{2dOLA_trade_off_xt005_mut000.png}), and we have chosen several values of $\alpha$ with which both the width of the averaging kernel $S$ and error magnification $V$ are sufficiently suppressed. After the selection of the candidates for a final value of $\alpha$, we looked at each of the candidate's averaging kernel, and we excluded values of $\alpha$ with which the localization of the averaging kernel is not well achieved. Eventually, we select one or two values of $\alpha$ and we obtain the corresponding estimates. The same procedures have been conducted for each target point. We describe the estimates thus obtained in the following subsections.

\begin{figure} [t]
\begin{center}
\includegraphics[scale=0.50]{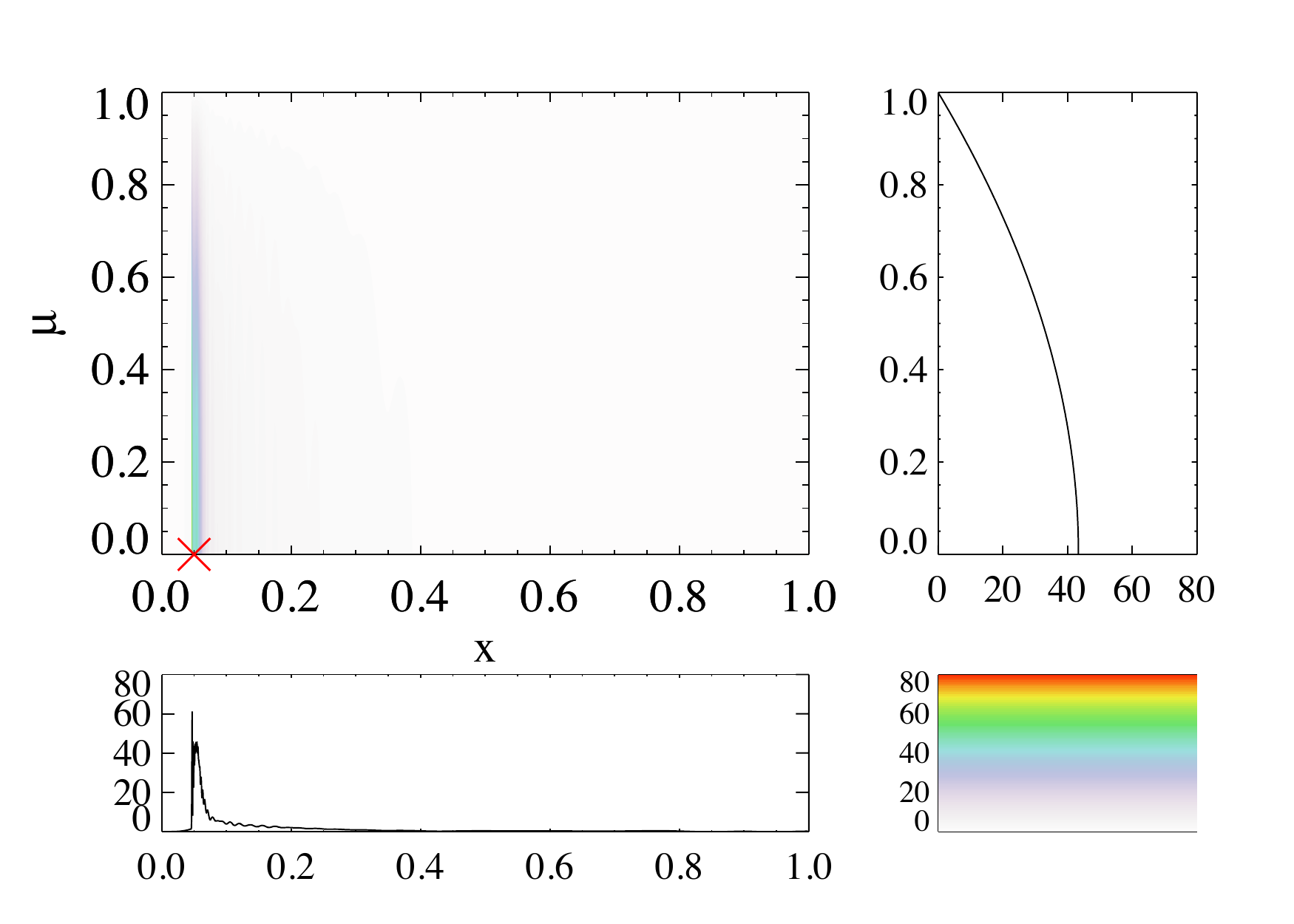}
\caption{\footnotesize Averaging kernel for the estimate at $(x,\mu)=(0.05,0.00)$, marked by the red cross, with $\alpha=10^{8}$ (the upper-left). The bottom-left and the upper-right figures are slices of the averaging kernel, taken at $\mu=0.00$ and $x=0.05$, respectively.}
\label{2DOLA_ave_krn_KIC11145123_alpha10_8_xt005_mt000.png}
\end{center} 
\end{figure}
\begin{figure} [hbt]
\begin{center}
\includegraphics[scale=0.5]{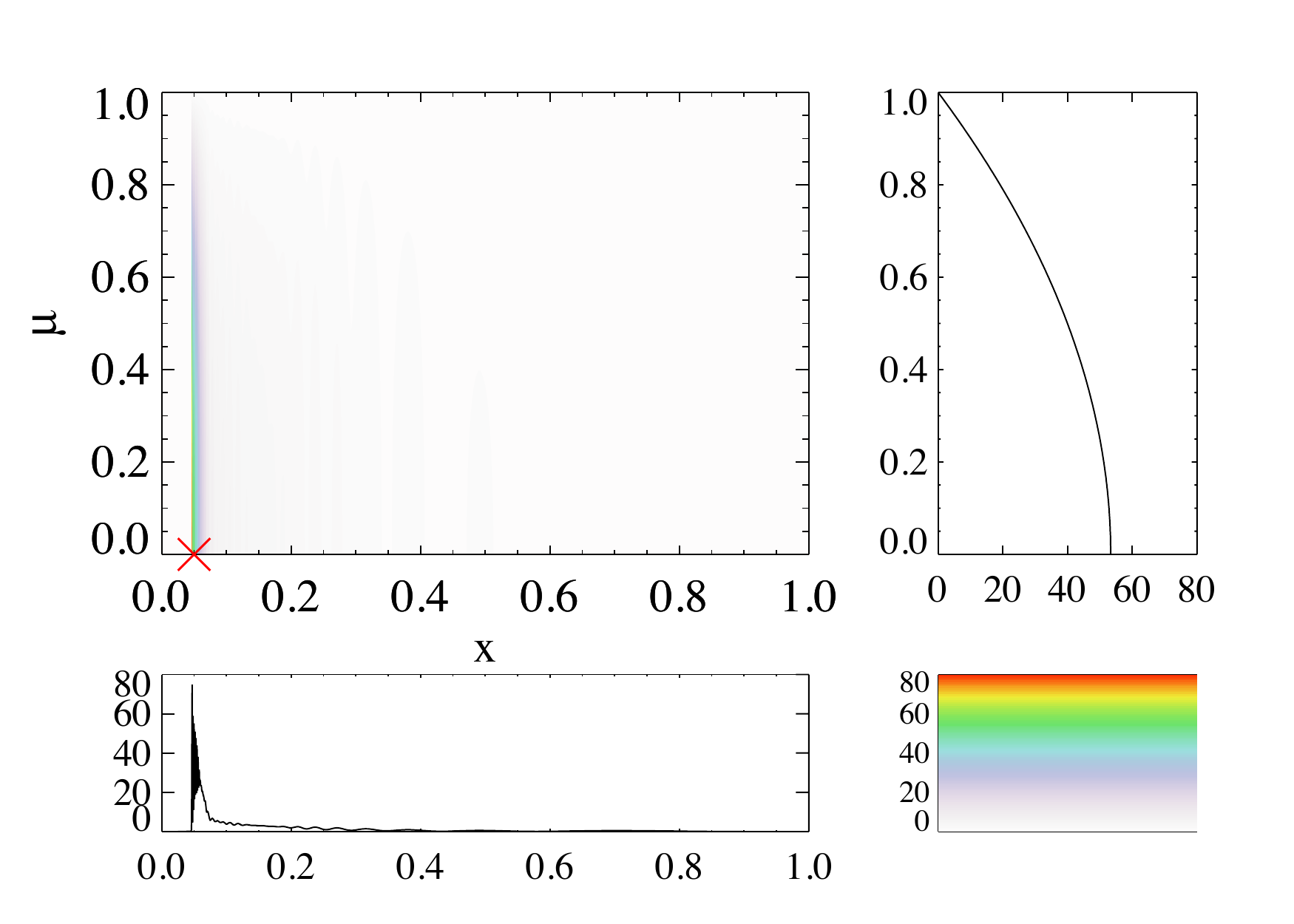}
\caption{\footnotesize Averaging kernel for the estimate at $(x,\mu)=(0.05,0.00)$, marked by the red cross, with $\alpha=10^{10}$ (the upper-left). The bottom-left and the upper-right figures are slices of the averaging kernel, taken at $\mu=0.00$ and $x=0.05$, respectively.}
\label{2DOLA_ave_krn_KIC11145123_alpha10_10_xt005_mt000.png}
\end{center} 
\end{figure}
\begin{figure} [hbt]
\begin{center}
\includegraphics[scale=0.5]{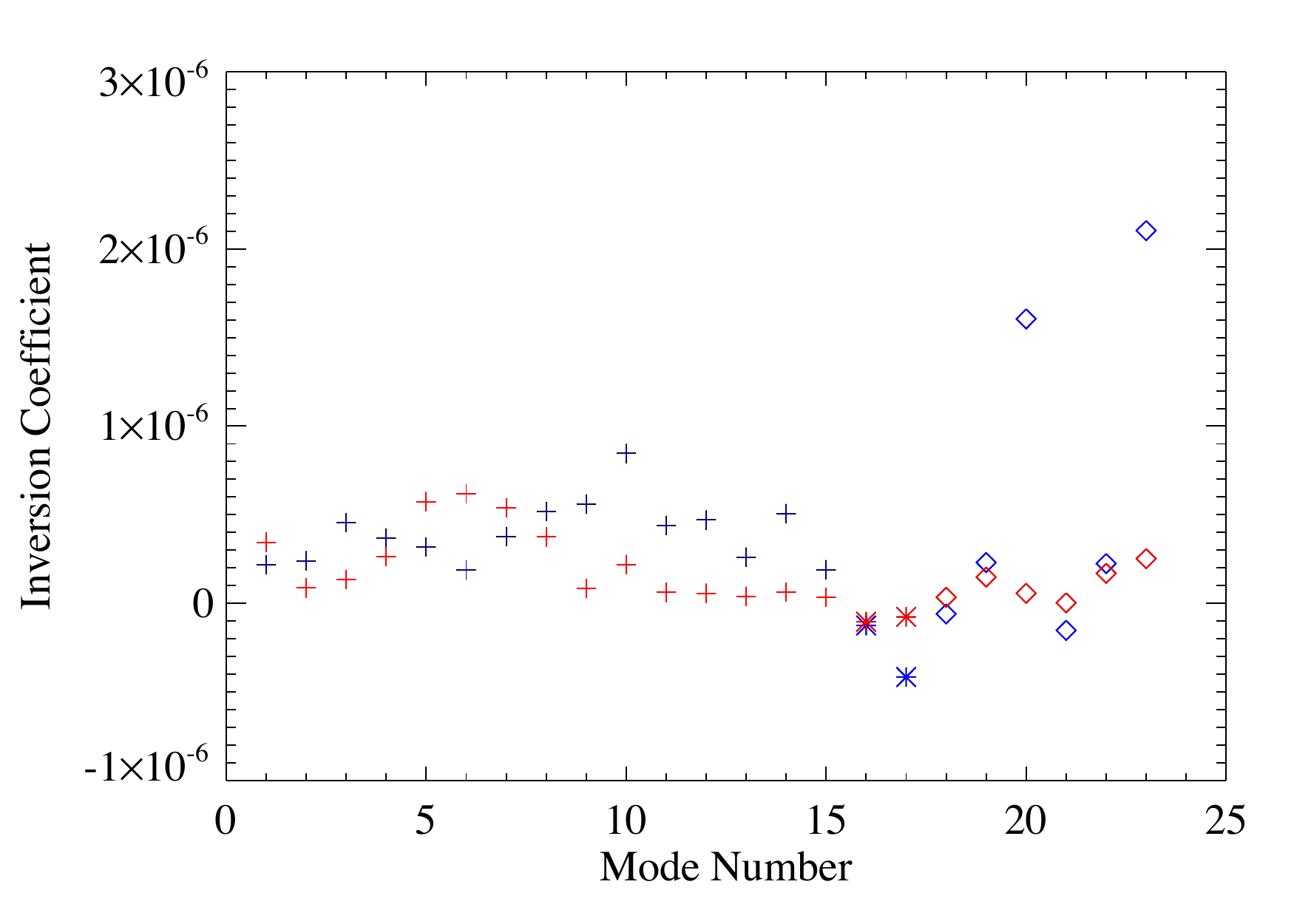}
\caption{\footnotesize Inversion coefficients that result from the OLA method with $\alpha=10^8$ (the blue marks) and $\alpha=10^{10}$ (the red marks). The target point is $(x,\mu)=(0.05,0.00)$. The crosses, the asterisks, and the diamonds represent the g modes, the p modes with $l=1$, and the mixed modes with $l=2$, respectively (for definition of the horizontal axis, mode number, see the caption of table \ref{freqs_modeled}). Note that our inversion coefficients are for rotational splittings, not for mode frequencies.}
\label{inv_coeff1}
\end{center} 
\end{figure}
\subsubsection{The estimate for the target point $(x,\mu)=(0.05,0.00)$}
\label{4-1-3}
For this target point, we present two equally reasonable estimates (\ref{estimate3-1}) and (\ref{estimate3-2}). The first estimate is obtained for $\alpha=10^8$, and the estimate and its standard deviation are:
\begin{equation}
\hat{\Omega}(0.05,0.00) = (0.9940\pm 0.0003)\,  \Omega_{100}, \label{estimate3-1}
\end{equation}
where $\Omega_{100}=2 \pi \times 0.01\, \rm{d}^{-1}$. The averaging kernel is localized well around the core region at low latitude (Figure \ref{2DOLA_ave_krn_KIC11145123_alpha10_8_xt005_mt000.png}), suggesting the high reliability of the estimate. The second estimate is obtained for $\alpha=10^{10}$, and the estimate and its standard deviation are:
\begin{equation}
\hat{\Omega}(0.05,0.00) = (0.9492\pm 0.0001)  \, \Omega_{100}. \label{estimate3-2}
\end{equation}
The averaging kernel is, again, localized well around the core region at low latitude (Figure \ref{2DOLA_ave_krn_KIC11145123_alpha10_10_xt005_mt000.png}).

Though we have achieved the localization of the averaging kernels reasonably well in both cases, there is a significant difference between the estimates (\ref{estimate3-1}) and (\ref{estimate3-2}). This difference can be attributed to the trade-off relation between resolution and error magnification. With the larger trade-off parameter $\alpha=10^{10}$, suppressing error magnification is emphasized (see equation (\ref{equation10})), leading to the smaller estimated standard deviation in the estimate (\ref{estimate3-2}) than that in the estimate (\ref{estimate3-1}). Meanwhile, localizing the averaging kernel is prioritized in the case of $\alpha=10^{8}$, and thus, the corresponding averaging kernel behaves better than that with $\alpha=10^{10}$; when we carefully look into the averaging kernel with $\alpha=10^{10}$, we find that it has an oscillatory component between $x=0.05$ and $x=0.075$ compared with that with $\alpha=10^{8}$. Therefore, we can qualitatively explain the behavior of the estimates. 

\begin{figure} [t]
\begin{center}
\includegraphics[scale=0.5]{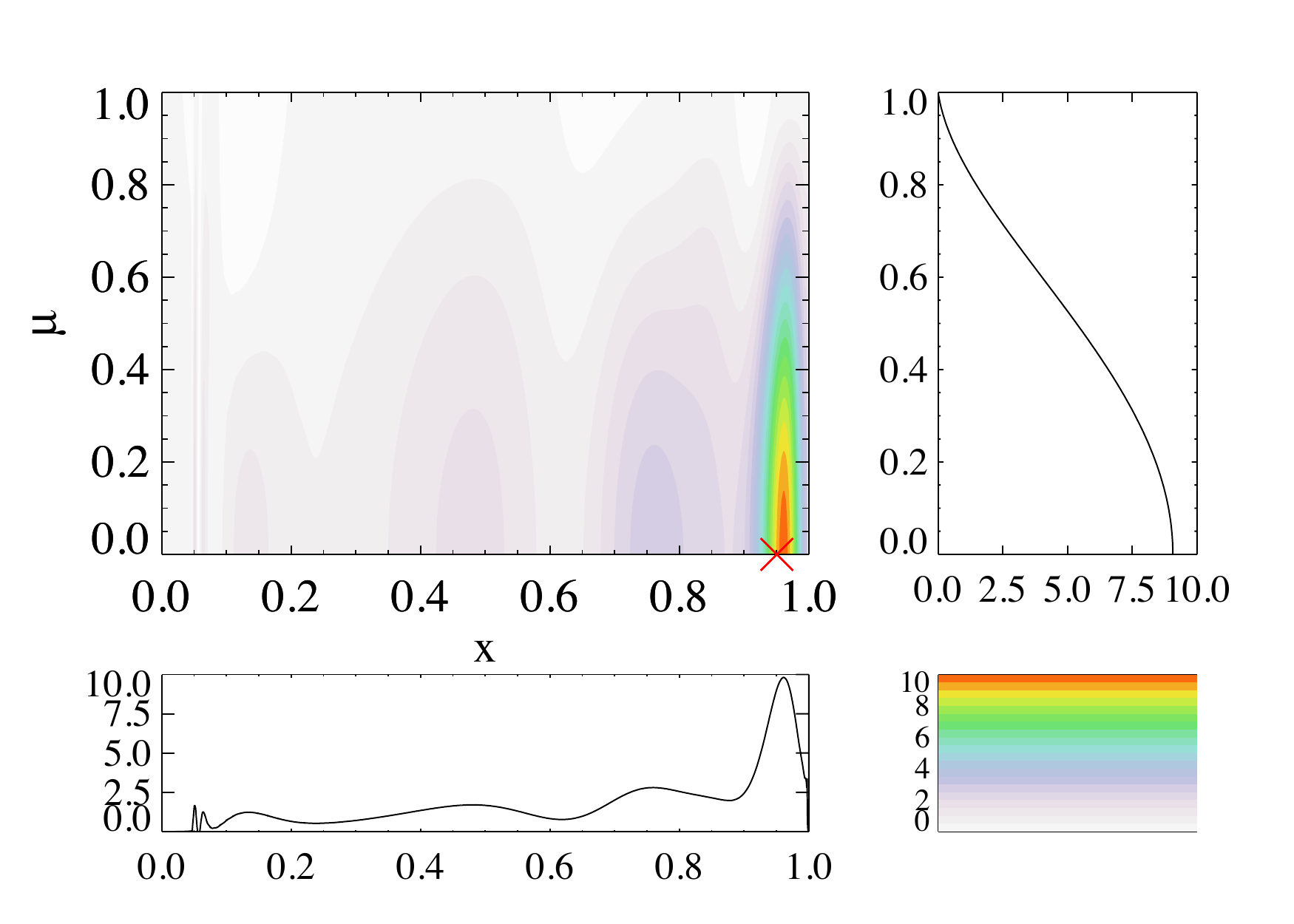}
\caption{\footnotesize Averaging kernel for the internal rotation estimate at $(x,\mu)=(0.95,0.00)$, marked by the red cross, with $\alpha=10^{8}$ (the upper-left). The bottom-left figure is a slice of the averaging kernel, at $\mu=0.00$, showing that the averaging kernel is distributed in the outer envelope. The upper-right figure is another slice of the averaging kernel, at $x=0.95$.}
\label{2DOLA_ave_krn_KIC11145123_alpha10_8_xt095_mt000.png}
\end{center} 
\end{figure}
For more quantitative hints, we take a look at inversion coefficients determined by rotation inversion with $\alpha=10^{8}$ and those with $\alpha=10^{10}$ (Figure \ref{inv_coeff1}). It is apparent that the averaging kernel for $\alpha=10^{10}$ is mainly composed of the g-mode splitting kernels. On the other hand, we have also found that the averaging kernel for $\alpha=10^8$ is composed not only of the g-mode splitting kernels but also of the mixed-mode splitting kernels. Thus, the distribution of the averaging kernels are slightly different. One possible explanation is as follows: there is a shear of the angular velocities around the core region, and thus, small differences between the two averaging kernels lead to the substantial difference in the two estimates. To test the possibility of such a fast-rotating core is one of the purposes in our three-zone modeling introduced in Section \ref{5-0}.

\begin{figure} [t]
\begin{center}
\includegraphics[scale=0.5]{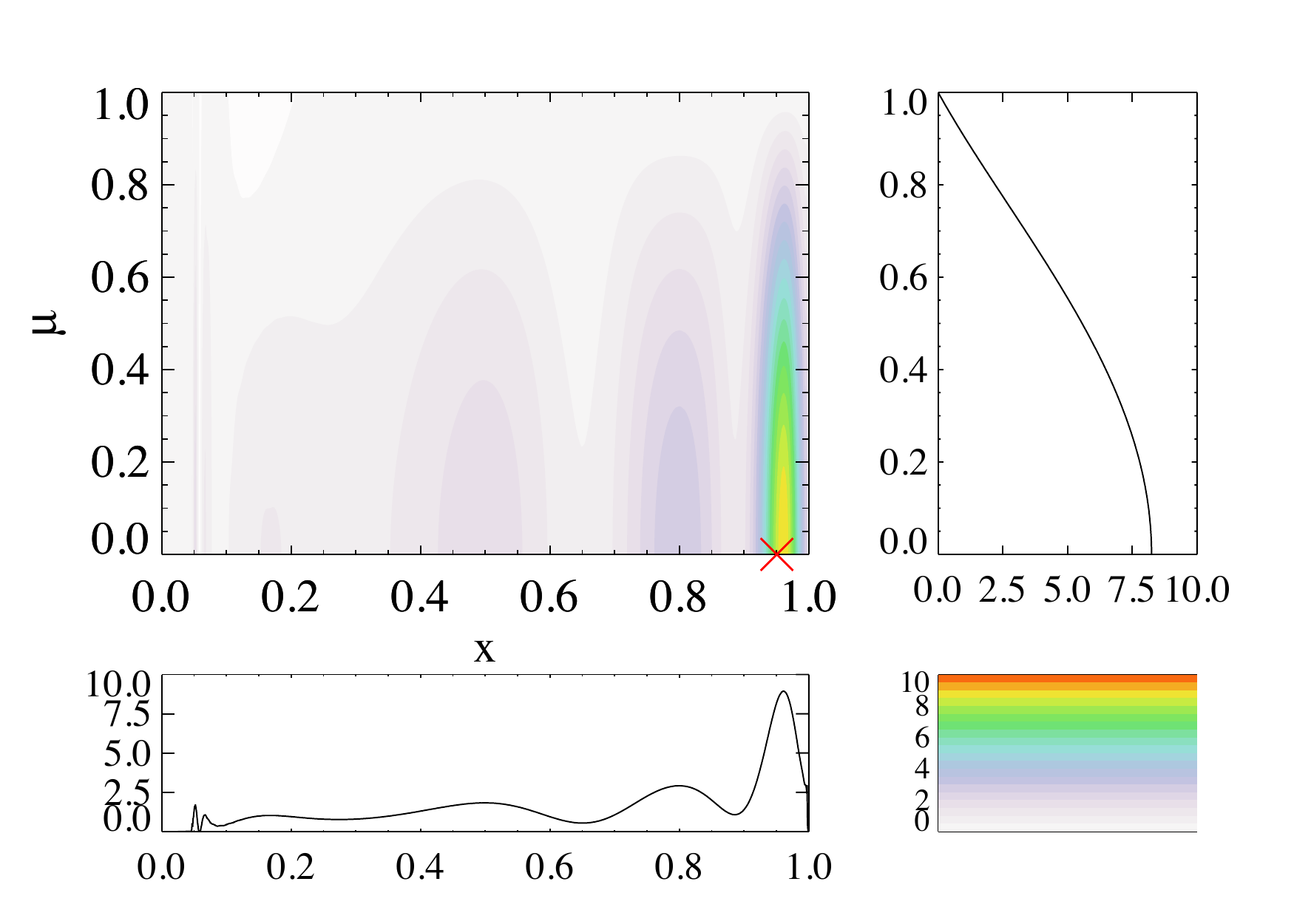}
\caption{\footnotesize Averaging kernel for the internal rotation estimate at $(x,\mu)=(0.95,0.00)$, marked by the red cross, with $\alpha=10^{10}$ (the upper-left). The bottom-left figure is a slice of the averaging kernel, at $\mu=0.00$, showing that the averaging kernel is distributed in the outer envelope. The upper-right figure is another slice of the averaging kernel, at $x=0.95$.}
\label{2DOLA_ave_krn_KIC11145123_alpha10_10_xt095_mt000.png}
\end{center} 
\end{figure}
\begin{figure} [t]
\begin{center}
\includegraphics[scale=0.5]{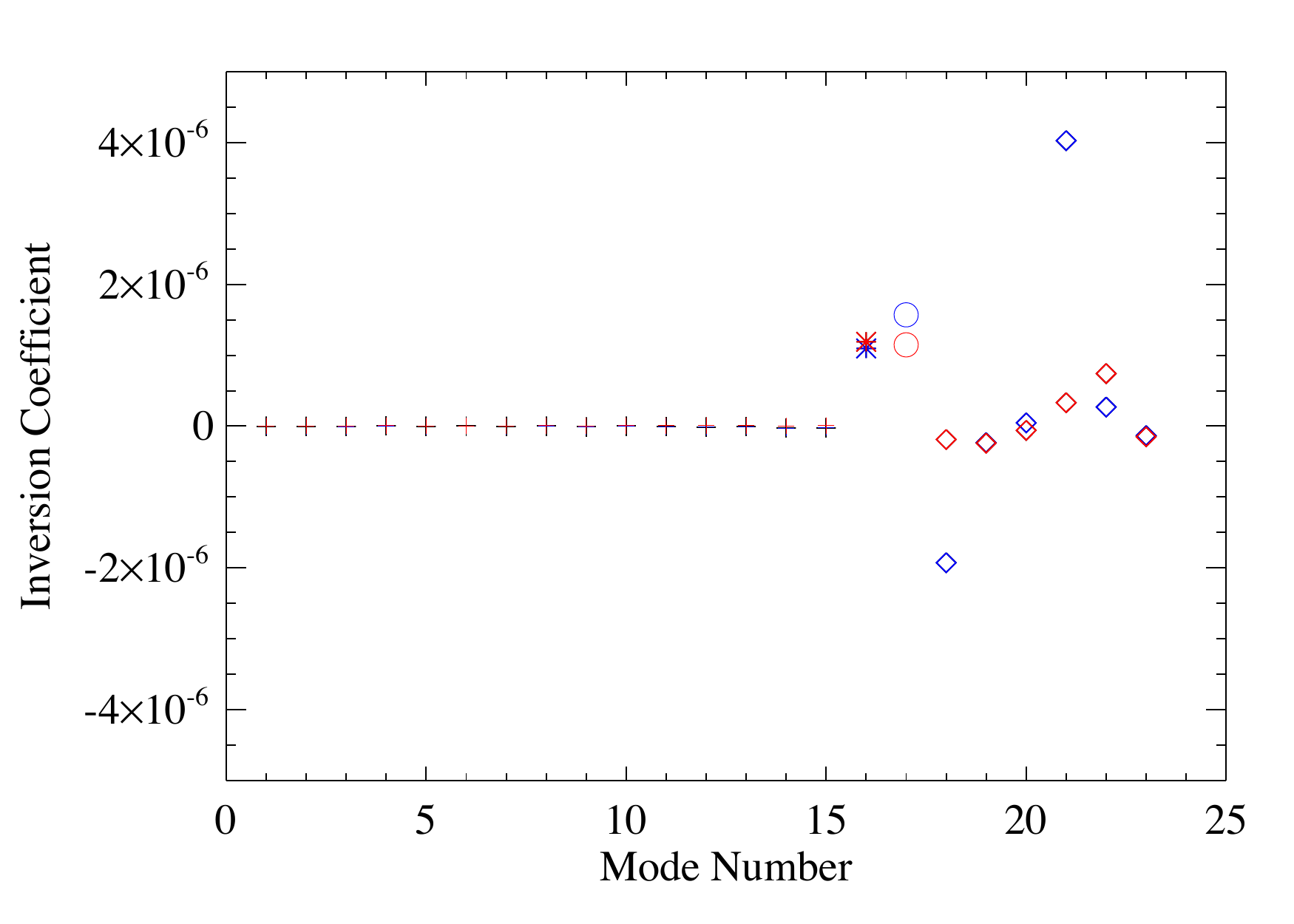}
\caption{\footnotesize Inversion coefficients resulting from the OLA method with $\alpha=10^8$ (the blue marks) and $\alpha=10^{10}$ (the red marks). The target point is $(x,\mu)=(0.95,0.00)$. The notation in this figure is the same as that in Figure \ref{inv_coeff1} except that for the mode with $(n,l)=(3,1)$, which is rather influential on estimation, the coefficients are indicated by open circles. }
\label{inv_coeff2}
\end{center} 
\end{figure}
\subsubsection{The estimate for the target point $(x,\mu)=(0.95,0.00)$}
\label{4-1-1}
We, again, show two equally reasonable estimates as follows: 
\begin{equation}
\hat{\Omega}(0.95,0.00) = (0.9511\pm 0.0005)\,  \Omega_{100}, \label{estimate1-1} 
\end{equation}
which has been obtained with $\alpha=10^{8}$, and
\begin{equation}
\hat{\Omega}(0.95,0.00) = (0.9706\pm 0.0002)\,  \Omega_{100}, \label{estimate1} 
\end{equation}
which has been obtained with $\alpha=10^{10}$. 

The averaging kernels for both estimates have peaks in the outer envelope (Figures \ref{2DOLA_ave_krn_KIC11145123_alpha10_8_xt095_mt000.png} and \ref{2DOLA_ave_krn_KIC11145123_alpha10_10_xt095_mt000.png}), and both of them are apparently localized around the target point. Nevertheless, we see that the averaging kernel for $\alpha=10^8$ has a higher peak than for $\alpha=10^{10}$ because more emphasis is put on resolving the averaging kernel in the case of $\alpha=10^8$. On the other hand, the estimate (\ref{estimate1-1}) has a larger standard deviation than the estimate (\ref{estimate1}) has since suppressing error magnification is less focused on in the case of $\alpha=10^8$ than in the case of $\alpha=10^{10}$. Thus, we clearly see the trade-off relation.

In order to see the origins of the significant difference in the two estimates, we assessed the inversion coefficients (Figure \ref{inv_coeff2}) as in the previous subsection. In the case of $\alpha=10^{8}$, the mixed-mode ($l=2$) splitting kernels are mainly used for the averaging kernel. However, in the case of $\alpha=10^{10}$, the p-mode ($l=1$) splitting kernels are mostly used for the averaging kernel. The above difference should lead to the two equally reasonable estimates. We would like to discuss in more detail the cause of the two reasonable estimates in Section \ref{5-095000}.

\begin{figure} [t]
\begin{center}
\includegraphics[scale=0.5]{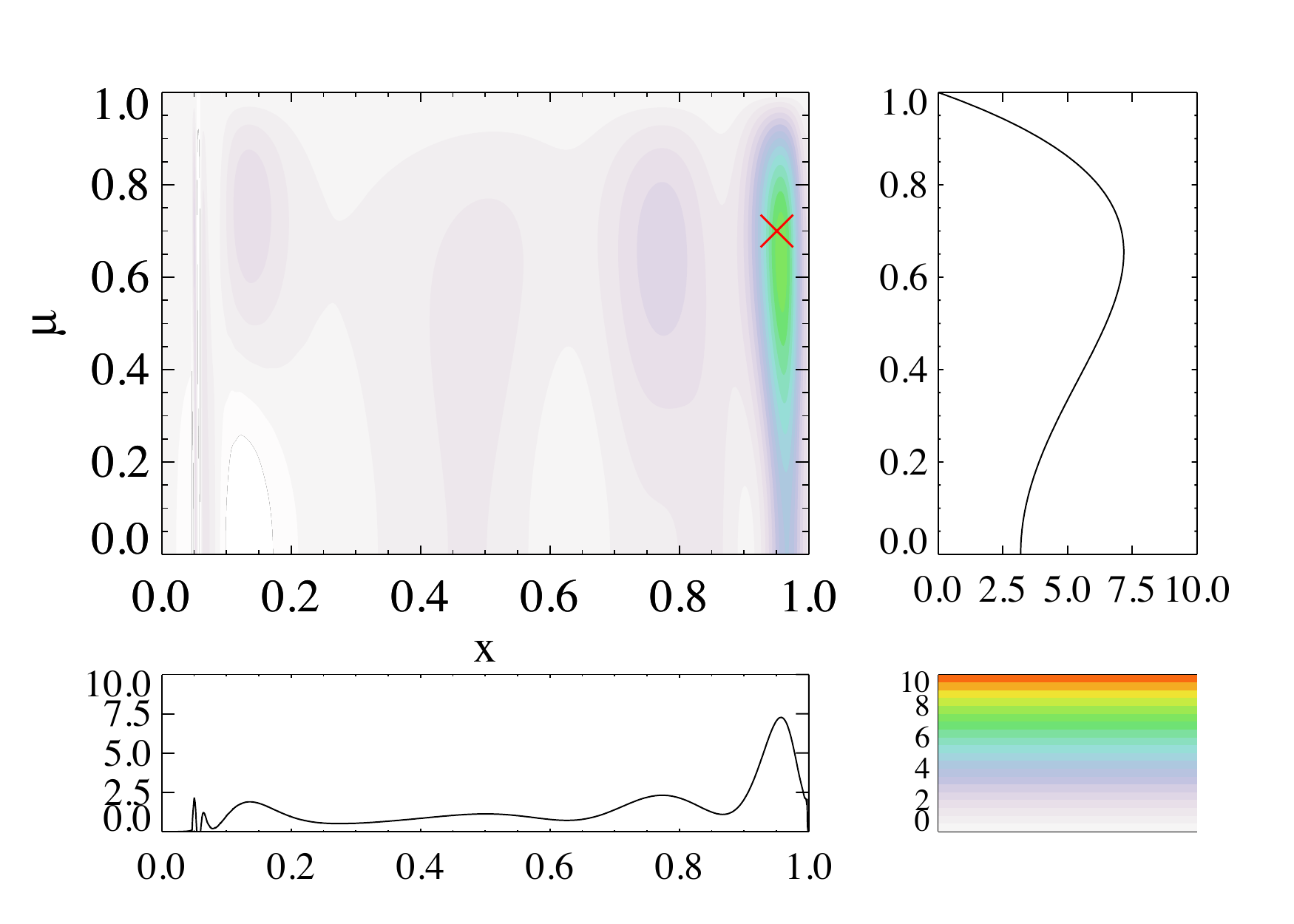}
\caption{\footnotesize Averaging kernel for the estimate at $(x,\mu)=(0.95,0.70)$, marked by the red cross, with $\alpha=10^{8}$ (the upper-left). The bottom-left figure is a slice of the averaging kernel, at $\mu=0.70$, showing that the averaging kernel is distributed in the outer envelope. The upper-right figure is another slice of the averaging kernel, at $x=0.95$ which indicates that the maximum of the averaging kernel is located in high-latitude region.}
\label{2DOLA_ave_krn_KIC11145123_alpha10_9_xt095_mt070.png}
\end{center} 
\end{figure}
\begin{figure} [hbt]
\begin{center}
\includegraphics[scale=0.5]{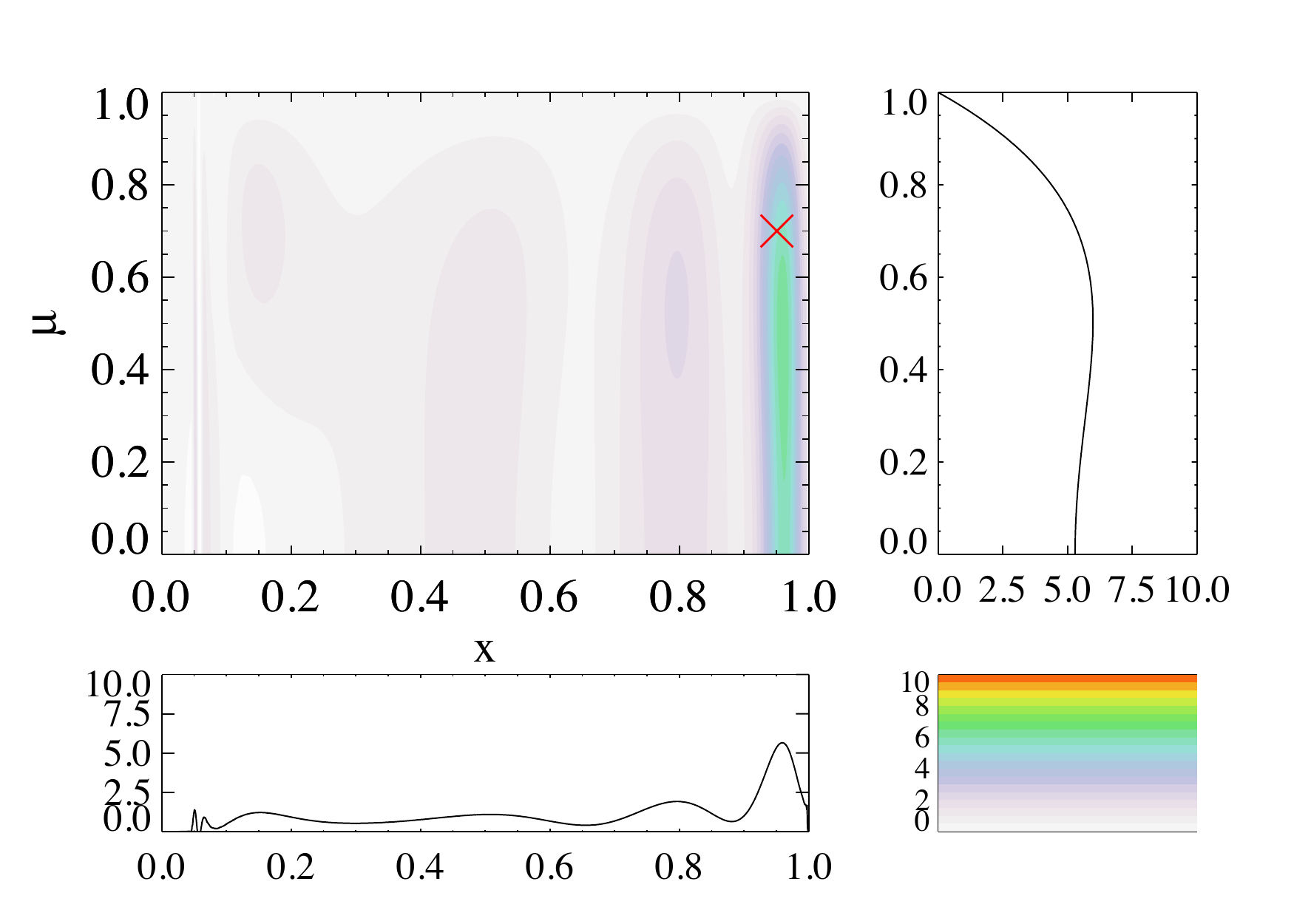}
\caption{\footnotesize Averaging kernel for the estimate at $(x,\mu)=(0.95,0.70)$, marked by the red cross, with $\alpha=10^{10}$ (the upper-left); more emphasis is put on decreasing the extents of error magnification than with $\alpha=10^{8}$. The bottom-left figure is a slice of the averaging kernel, at $\mu=0.70$. The upper-right figure is another slice of the averaging kernel, at $x=0.95$ which shows that the averaging kernel is broadly distributed and it is not well localized around the target point.}
\label{2DOLA_ave_krn_KIC11145123_alpha10_11_xt095_mt070.png}
\end{center} 
\end{figure}
\subsubsection{The estimate for the target point $(x,\mu)=(0.95,0.70)$}
\label{4-1-2}
Here, one estimate is obtained with $\alpha=10^8$ as follows: 
\begin{equation}
\hat{\Omega}(0.95,0.70) = (0.9564\pm 0.0006) \, \Omega_{100}. \label{estimate2} 
\end{equation}
The weak dependence of the estimate on the trade-off parameter $\alpha$ is due to the small number of splitting kernels which have sensitivity in high-latitude regions; we do not have any other choice. When we increased the importance of error magnification by increasing $\alpha$, we have failed in localizing the averaging kernel (Figure \ref{2DOLA_ave_krn_KIC11145123_alpha10_11_xt095_mt070.png}) in contrast to the case with the smaller value of $\alpha$ (Figure \ref{2DOLA_ave_krn_KIC11145123_alpha10_9_xt095_mt070.png}). By comparing the estimate (\ref{estimate2}) with the estimates (\ref{estimate1-1}) or (\ref{estimate1}), we find the possibility that there is latitudinally differential rotation in the outer envelope. We would like to expand our discussion on the existence of the latitudinally differential rotation in Section \ref{5-lat}.

\section{Discussion}
\label{5}
\subsection{Three-zone modeling}
\label{5-0}
We have carried out the three-zone modeling in order to test the possibility of a velocity shear around the core region (see Subsection \ref{4-1-3}), which has been suggested from the results obtained by the OLA method. In our three-zone modeling, we assume a constant angular velocity for each of three regions, namely, the innermost region $(0<x<x_a,0<\mu<1)$: region 1, the inner region $(x_a<x<x_b,0<\mu<1)$: region 2, and the outer region $(x_b<x<1,0<\mu<1)$: region 3. The positions of boundaries, denoted by $x_{a}$ and $x_{b}$, are treated as free parameters. 

The linear integral equation (\ref{equation5}) can be expressed in a much simpler form as below: 
\begin{equation}
d_{i} = \sum_{j=1}^{3} K_{ij}\Omega_{j} + e_{i}, \; \; \; \; i=1,...,M,  \; \; \; \;  j=1,2,3 \label{equation4-2-1}
\end{equation}
where $K_{ij}$ is defined as
\begin{equation}
K_{ij} \equiv  \int \!\!\! \int_{j} K_{i}(x,\mu) dx d \mu, \label{equation4-2-2}
\end{equation}
and the integration is carried out over the region $j$. 

Equation (\ref{equation4-2-1}) can be rewritten in the following form
\begin{equation}
\bm{d}=\mathbf{K} \bm{\Omega} + \bm{e}, \label{equation4-2-3}
\end{equation}
where $\mathbf{K}$ is a $23 \times 3$ matrix. The data, the angular velocities, and the errors are denoted by $\bm{d}$, $\bm{\Omega}$, and $\bm{e}$, respectively. The linear inverse problem (\ref{equation4-2-3}) was relatively easy to solve in a least-squares sense because the rank of the observation matrix $\mathbf{K}$ is found to be full. We have repeated the inversion changing the values of the three free parameters in the ranges as follows: $0.034<x_a<0.048$ and $0.19<x_b<0.95$. 

\begin{figure} [t]
\begin{center}
\includegraphics[scale=0.18]{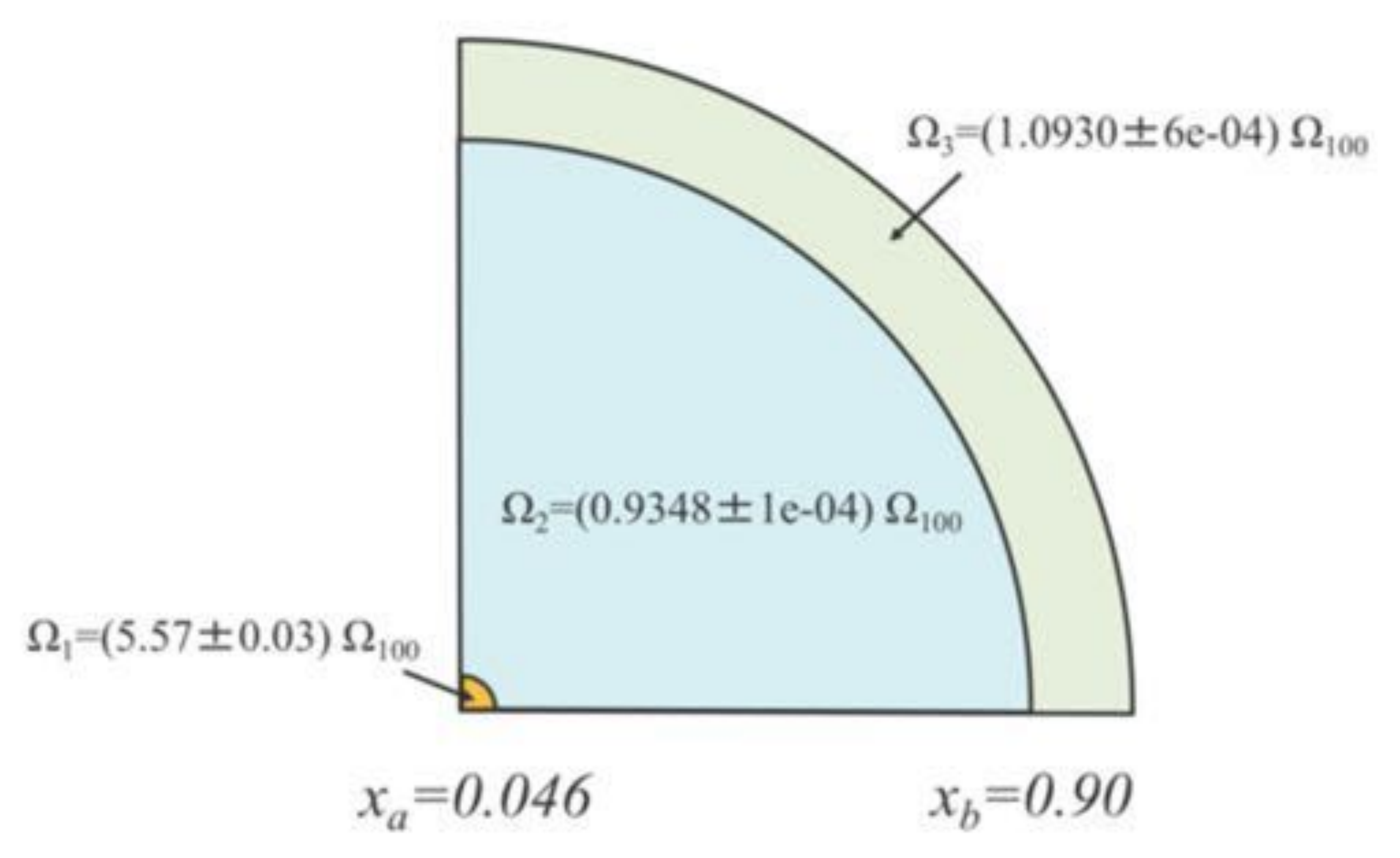}
\caption{\footnotesize Estimates obtained by the three-zone modeling of the internal rotation. The unit $\Omega_{100}$ is defined in the text. The innermost region, which is identical to the convective core, rotates much faster than the other parts. We also confirm the same trend obtained by \citet{Kurtz2014} in the outer envelope; the inner region rotates slightly slower than the outer region does.}
\label{three_zone_modeling.png}
\end{center} 
\end{figure}

The estimates of the internal rotation which makes the residual of the inversion minimum is as follows:
\begin{equation}
\Omega_{1} = (5.57\pm 0.03)\,  \Omega_{100}, \label{3_zone_1} 
\end{equation}
\begin{equation}
\Omega_{2} = (0.9348\pm 0.0001)\,  \Omega_{100}, \label{3_zone_2} 
\end{equation}
\begin{equation}
\Omega_{3} = (1.0930\pm 0.0006)\,  \Omega_{100}, \label{3_zone_3} 
\end{equation}
which are obtained when $x_a$ is 0.046 and $x_b$ is 0.905 (see Figure \ref{three_zone_modeling.png}). Interestingly, we have obtained a result which suggests that the innermost region rotates about $6$ times faster than the other parts of the star. Moreover, the inner boundary $x_a$ is almost identical to the edge of the convective core. This result indeed indicates the existence of the shear of the angular velocities, as suggested in the last section. We also see that the deep radiative region (region 2) rotates slightly slower than the outer region (region 3), which is the same trend as indicated by the result of the two-zone modeling in \citet{Kurtz2014}.

\subsection{An explanation for the two possible estimates for the central target point}
\label{5-1}
The apparent discrepancy between the estimates (\ref{estimate3-1}) and (\ref{estimate3-2}) can be fully explained provided that the fast rotating core inferred from the three-zone modeling is valid. The key to understand the discrepancy is the averaging kernel for each case. Figure (\ref{inv_coeff1}) shows the inversion coefficients $c_{i}$ for each case ($\alpha=10^8$ and $\alpha=10^{10}$), from which we see the contribution of each splitting kernel to the corresponding averaging kernel. As we have already mentioned in Section \ref{4-1}, the distributions of the averaging kernels are slightly different.

\begin{figure} [t]
\begin{center}
\includegraphics[scale=0.5]{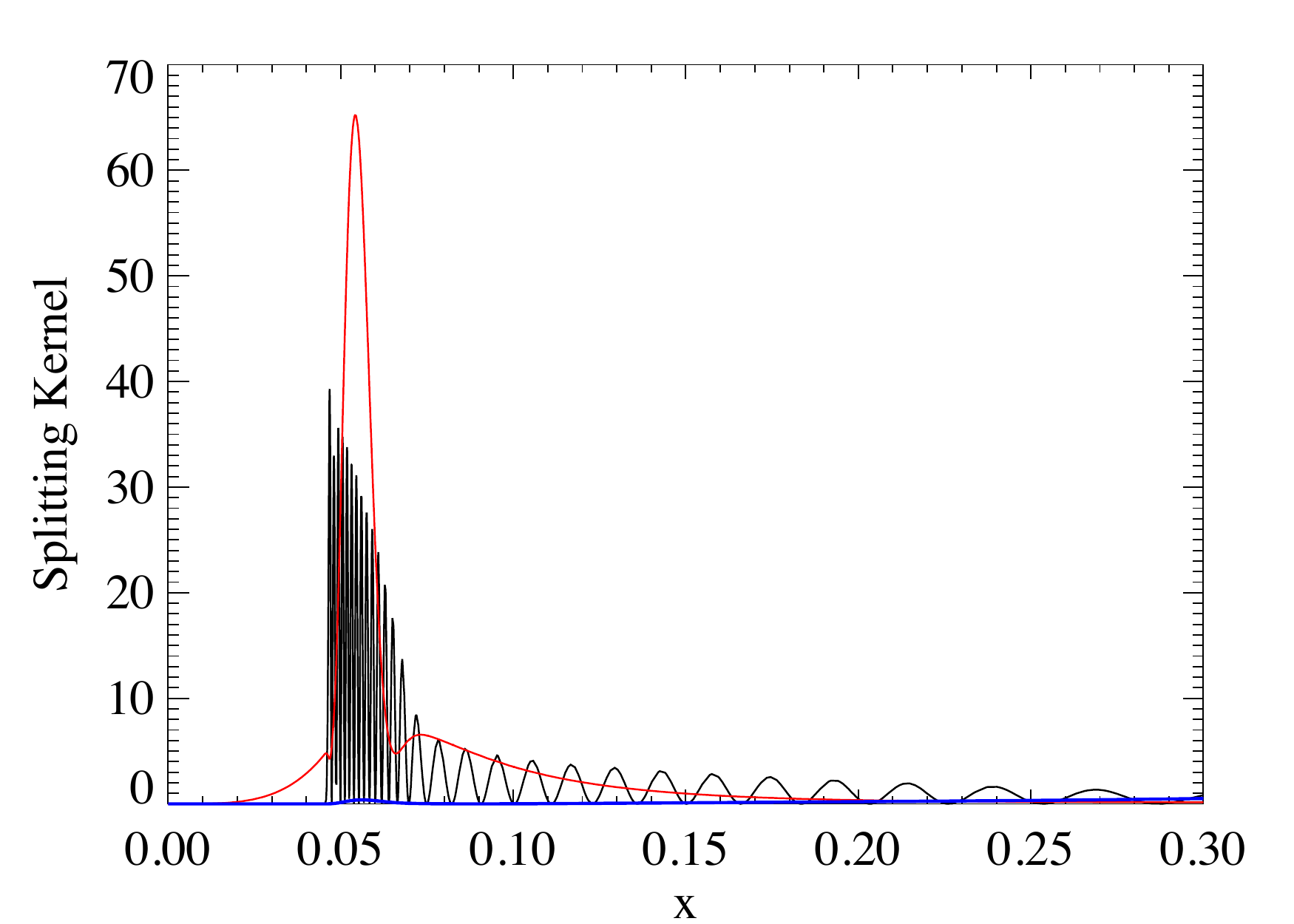}
\caption{\footnotesize Expanded look in the slices of the splitting kernels at $\mu=0.0$. The black curve is the splitting kernel of the g mode $(n,l,m)=(-33,1,1)$ and it does not have any sensitivity inside the convective core below $x=0.045$. The blue curve is the splitting kernel of the p mode $(n,l,m)=(2,1,1)$ and it exhibits almost no sensitivity in this deep region (almost identical to the horizontal axis). Finally, the red curve is the splitting kernel of one of the other mixed modes, with $(n,l,m)=(2,2,2)$. It evidently has sensitivity below $x=0.045$ which is the boundary between the convective core and the radiative region. The difference in the sensitivity between the mixed mode with $(2,1,1)$ and that with $(2,2,2)$ is caused by the fact that the modes behave differently inside the star; the former behaves as a p mode in most part of the star, and in contrast, the latter has a g-mode nature around the core region, leading to such a high sensitivity.}
\label{slices_of_kernels}
\end{center} 
\end{figure}

When we compare slices of the g-mode splitting kernels with those of the mixed modes, we find a difference between them (Figure \ref{slices_of_kernels}), particularly for the mixed mode with $(n,l)=(2,2)$. The g-mode splitting kernels do not have sensitivity inside the convective core since g modes are not propagative there. However, the mixed-mode splitting kernel with $(n,l)=(2,2)$ has, albeit rather small, sensitivity well below the edge of the convective core ($x=0.045$) which has been inferred to rotate about 6 times faster than the other parts of the star. 

The discussion in the previous paragraph enables us to explain the reason why there exist two possible estimates. In the case of $\alpha=10^{10}$, the g-mode splitting kernels are mainly used to obtain the averaging kernel, and because they do not have sensitivity inside the convective core, there is no contribution of the fast rotating core to the slower estimate (\ref{estimate3-2}). On the other hand, in the case of $\alpha=10^8$, the mixed-mode splitting kernel with $(n,l)=(2,2)$ is used, and because it has sensitivity inside the convective core, there is a contribution of the fast rotating core to the faster estimate (\ref{estimate3-1}). Accordingly, we have selected the estimates (\ref{estimate3-2}) as a final estimate of the internal rotation of the deep radiative region of the star. It is thus evident that the deep radiative region rotates slower than the envelope does because the estimate (\ref{estimate3-2}) is significantly smaller than the estimates (\ref{estimate1-1}), (\ref{estimate1}), and (\ref{estimate2}).

%
\subsection{An explanation for the two possible estimates for the target point $(x,\mu)=(0.95,0.00)$}
\label{5-095000}
In this section, we attempt to explain the origin of the seemingly contradictory estimates (\ref{estimate1-1}) and (\ref{estimate1}). We first look at the inversion coefficients of rotation inversion with the target point $(x,\mu)=(0.95,0.00)$ (Figure \ref{inv_coeff2}). As we have already mentioned in Section \ref{4-1}, the p- or mixed-mode splitting kernels are mainly used for the averaging kernels in both cases ($\alpha=10^{8}$ and $\alpha=10^{10}$). Then, we repeated rotation inversion excluding one certain p- or mixed-mode rotational shift after another from the inversion procedure to figure out which mode is the most influential on estimation. What we found is that estimates which are hardly dependent on the trade-off parameter $\alpha$ can be obtained when we exclude the p mode with $(n,l)=(3,1)$. 

\begin{figure} [t]
\begin{center}
\includegraphics[scale=0.5]{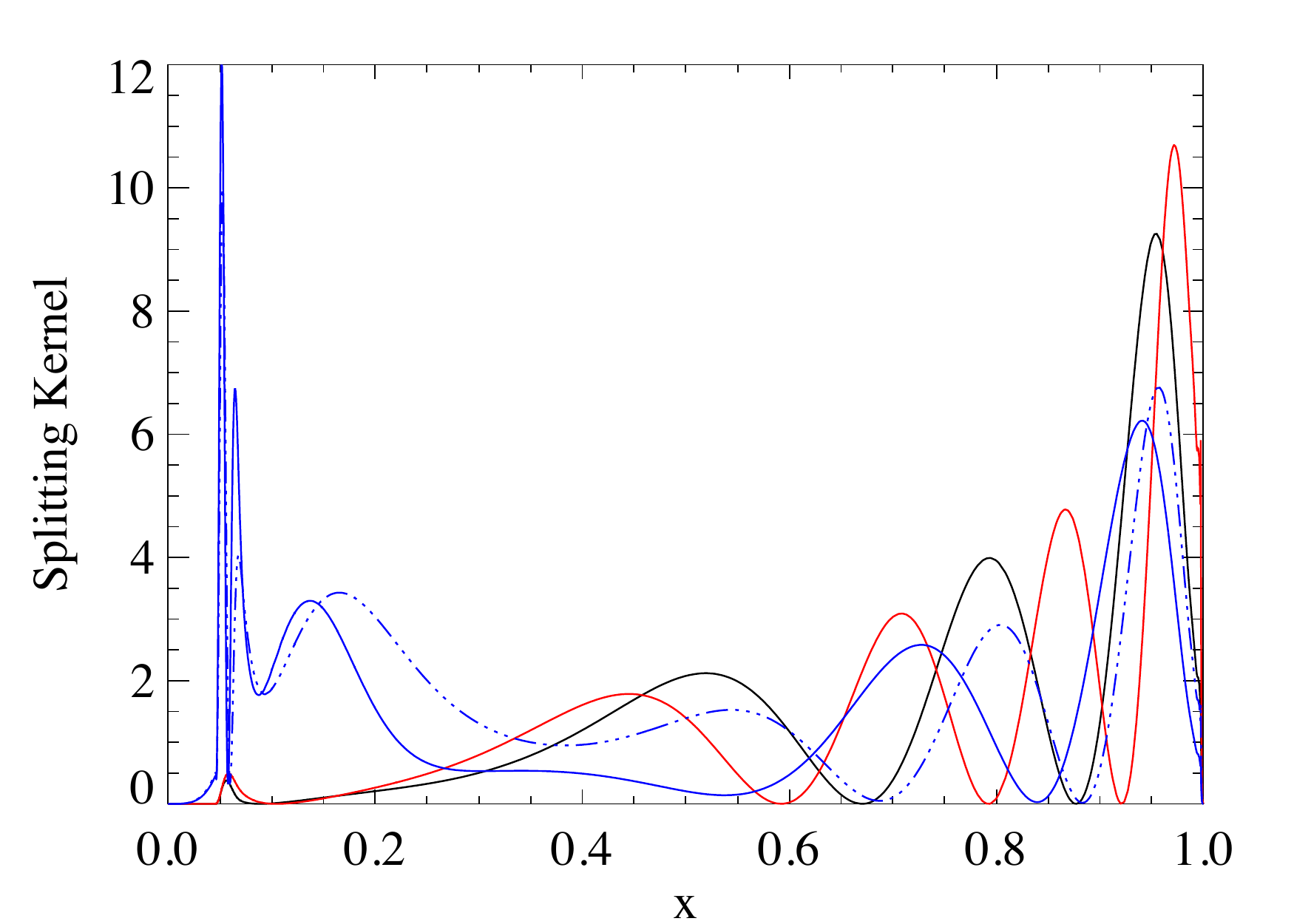}
\caption{\footnotesize Slices of the splitting kernels at $\mu=0.0$. The red and black curves are representing the splitting kernels of the p modes with $(n,l,m)=(2,1,1)$ and $(3,1,1)$, respectively. The mixed-mode splitting kernels are also shown as the blue curves; the solid one is for the mode with $(n,l,m)=(-1,2,2)$, and the dot-dashed one is for the mode with $(n,l,m)=(0,2,2)$. We see that only the red curve has large sensitivity to internal rotation of the region $0.83<x<0.9$.}
\label{slices_of_kernels_2}
\end{center} 
\end{figure}

It is reasonable because the rotational shift of the mode is much smaller $(\sim0.0085 \ \rm{d}^{-1})$ than the other p-mode rotational shift $(\sim0.0097 \ \rm{d}^{-1})$, and the splitting kernel of the mode with $(n,l)=(3,1)$ has the highest peak near the target point $x=0.95$ (Figure \ref{slices_of_kernels_2}). Therefore, to increase the resolution of the averaging kernel which targets at $(x,\mu)=(0.95,0.00)$, the splitting kernel with $(n,l)=(3,1)$ is likely to be used, leading to smaller estimates such as the estimate (\ref{estimate1-1}).

However, since the radial structures of p- or mixed-mode splitting kernels (Figure \ref{slices_of_kernels_2}) are rather complex compared with those of the g modes (Figure \ref{slices_of_kernels}), it is difficult to describe the reason why the mode with $(n,l)=(3,1)$ is so influential on the estimation; the situation is not as simple as that we discuss in Section \ref{5-1}. We find an implication that the outer region between $0.83<x<0.9$, where the splitting kernel with $(n,l)=(3,1)$ has greater sensitivity than the other kernels have, might rotate much slower than the other outer regions. This is because the latitudinal dependence of the mode with $(n,l)=(3,1)$ is completely the same as that of the mode with $(n,l)=(2,1)$, and thus, the difference in the rotational shifts must derive from the difference in radial sensitivity of each splitting kernel. Note that we cannot decide whether the estimate (\ref{estimate1-1}) is better than the other estimate (\ref{estimate1}) or not due to the difficulty in interpreting the rotational shift of the mode with $(n,l)=(3,1)$.

\subsection{The evidence of latitudinally differential rotation}
\label{5-lat}
So far, we have not been able to select one estimate from the two equally reasonable estimates (\ref{estimate1-1}) and (\ref{estimate1}) for the target point $(x,\mu)=(0.95,0.00)$. Thus, we cannot tell whether the high-latitude region rotates faster than the low-latitude region does or not because the estimates are ordered as: (\ref{estimate1-1}) $<$ (\ref{estimate2}) $<$ (\ref{estimate1}). Actually, we probably obtain innumerable estimates by changing the value of $\alpha$ between $10^8$ and $10^{10}$, and thus, it is possible that there exist estimates based on which we do not find any latitudinal dependence of the internal rotation in the outer envelope.

In order to check whether latitudinally differential rotation does exist or not, we focus on the three quintuplets which are identified as $l=2$ (see Table \ref{freqs_modeled}) and we directly compute so-called ``a-coefficients'' of the three multiplets. The a-coefficients are coefficients which are used in an expanded expression of a frequency shift as follows:
\begin{equation}
 \omega_{nlm} - \omega_{nl0}  = \sum_{k} a_{k}(n,l) \mathcal{P}_{k}(m;l), \label{equation5-lat-1}
\end{equation}
where $k$ is the order of the expanding polynomial $\mathcal{P}_{k}(m;l)$ \citep{Ritzwoller1991}. In the present paper, we followed the formulation of \citet{Schou1994}. It is easily shown that the odd-order terms $a_{2k+1}$ correspond to rotational splitting by substituting the expression (\ref{equation5-lat-1}) into the definition (\ref{equation0}). In particular, the sign of the $a_3$ coefficient represents the latitudinal dependence of the internal rotation; the positive (negative) value corresponds to the faster internal rotation in the low (high) latitude region. From the observed rotational shifts, we computed three $a_{3}$ coefficients for the quintuplets $(n,l)=(-1,2), (0,2), (2,2)$ with the propagated standard deviations as below:
\begin{displaymath}
a_{3}^{\rm{obs}}(-1,2)=(-6.7\pm 3.3) \times 10^{-6} \  \rm{d}^{-1}  \label{equation5-lat-2}
\end{displaymath}
\begin{displaymath}
a_{3}^{\rm{obs}}(0,2)=(0.15\pm 1.3) \times 10^{-6}  \  \rm{d}^{-1}  \label{equation5-lat-3}
\end{displaymath}
\begin{displaymath}
 a_{3}^{\rm{obs}}(2,2)=(-1.8\pm 8.6) \times 10^{-6} \  \rm{d}^{-1}.  \label{equation5-lat-4}
\end{displaymath}

As we see, $a_{3}^{\rm{obs}}(-1,2)$ is smaller than zero with more than $2\sigma$ significance, suggesting that the high-latitude region rotates faster than the low-latitude region does. Nevertheless, both $a_{3}^{\rm{obs}}(0,2)$ and $a_{3}^{\rm{obs}}(2,2)$ are zero within the error bars. For assessing whether there is latitudinal dependence of the internal rotation or not based on all the three $a_3$ coefficients, we carried out a simple statistical test. The null hypothesis we would like to reject is that $a_3(n,l)=0$ for any multiplet, and we assume that the probability density function of each observed $a_{3}$ coefficient is distributed as Gaussian with the mean and the standard deviation equal to zero and its observational estimate, respectively. Then, we calculated probability that we measure $a_3$ coefficient whose absolute value is larger than that of the actually observed one, $a_{3}^{\rm{obs}}(\it{n},l)$, as below:
\begin{displaymath}
p(n,l)=\int_{\rm{out}} \frac{1}{\sqrt{2\pi} \sigma(n,l)} \rm{exp} \biggl( -\frac{1}{2} \frac{(\textit{x}-0)^2}{\sigma(\textit{n},\textit{l})^2} \biggr) \textit{dx},  \label{equation5-lat-5}
\end{displaymath}
where $x$ is a dummy variable and $\sigma(n,l)$ is an observational estimate of the standard deviation of $a_{3}^{\rm{obs}}(\it{n},l)$. The integration is carried out over the region, $-\infty < x < -a_{3}^{\rm{obs}}(\it{n},l)$ or $a_{3}^{\rm{obs}}(\it{n},l) < x < \infty$. Note that $a_{3}^{\rm{obs}}(\it{n},l)$ represents the actual observed value here. Thus, the probability that we measure a set of $a_3$ coefficients $\{a_{3}(-1,2), a_{3}(0,2), a_{3}(2,2)\}$ which satisfies the following conditions 
\begin{eqnarray}
&|&a_{3}(-1,2)| \geq |a_{3}^{\rm{obs}}(-1,2)| \nonumber  \\
&|&a_{3}(0,2)| \geq |a_{3}^{\rm{obs}}(0,2)| \nonumber \\
&|&a_{3}(2,2)| \geq |a_{3}^{\rm{obs}}(2,2)| \nonumber 
\end{eqnarray}
are simply calculated as below:
\begin{displaymath}
p = p(-1,2) \times p(0,2) \times p(2,2) ,  \label{equation5-lat-7}
\end{displaymath}
and we found that the value of $p$ is 0.033. This is more than $2\sigma$ significance but less than $3\sigma$ significance, and thus, we conclude that it is marginal to reject the null hypothesis and to claim that $a_3 \neq 0$. Nevertheless, if we admit the existence of the latitudinally differential rotation, it is implied that the high-latitude region rotates faster than the low-latitude region does since the observed $a_3$ coefficients are mostly negative. This feature is sometimes called anti-solar differential rotation \citep[e.g.][]{Brun2017}. Therefore, we consider the estimate (\ref{estimate1-1}) as a better choice than the estimate (\ref{estimate1}); the former estimate is shown to be supported by the less model-dependent inference.

\subsection{Is the fast spinning core reasonable?}
\label{5-2}
\citet{Aerts2017} summarized the studies on the internal rotation of main-sequence A-F type stars, and they concluded that almost all the stars investigated so far exhibited nearly rigid rotation. This is different from the result we have obtained through the present study; there can be a strong velocity shear between the convective core and the radiative region above. Our result, however, is actually compatible with their view because all the studies in \citet{Aerts2017} have estimated the ``core'' rotation based on g modes with which we cannot extract the information on the convective core. Thus, what they have inferred are internal rotation of the ``deep radiative'' region above the convective core. Meanwhile, our estimation of ``core'' rotation is based on a mixed mode which has sensitivity inside the convective core. We have succeeded in extracting information on the convective core, which enables us to reveal the fast-core rotation of KIC11145123. Furthermore, our results show that the radiative region of the star rotates nearly rigidly (e.g. see the estimates \ref{3_zone_2} and \ref{3_zone_3}). This result is consistent with the current understanding of the internal rotation of A-F stars.

From the theoretical point of view, there have been a series of numerical simulations of the dynamo mechanism inside the convective core of A type stars \citep[e.g.][]{Browning2004, Brun2005, Featherstone2009} where internal differential rotation has been also calculated. In particular, though they focused more on the magnetism of A stars, one of their studies \citep{Browning2004} reproduced internal rotation profile in which the convective core rotates a few times faster than the radiative region above, which is similar to what we have found in this study. The fast-core rotation in an A star might be theoretically feasible. 

\begin{figure} [t]
\begin{center}
\includegraphics[scale=0.5]{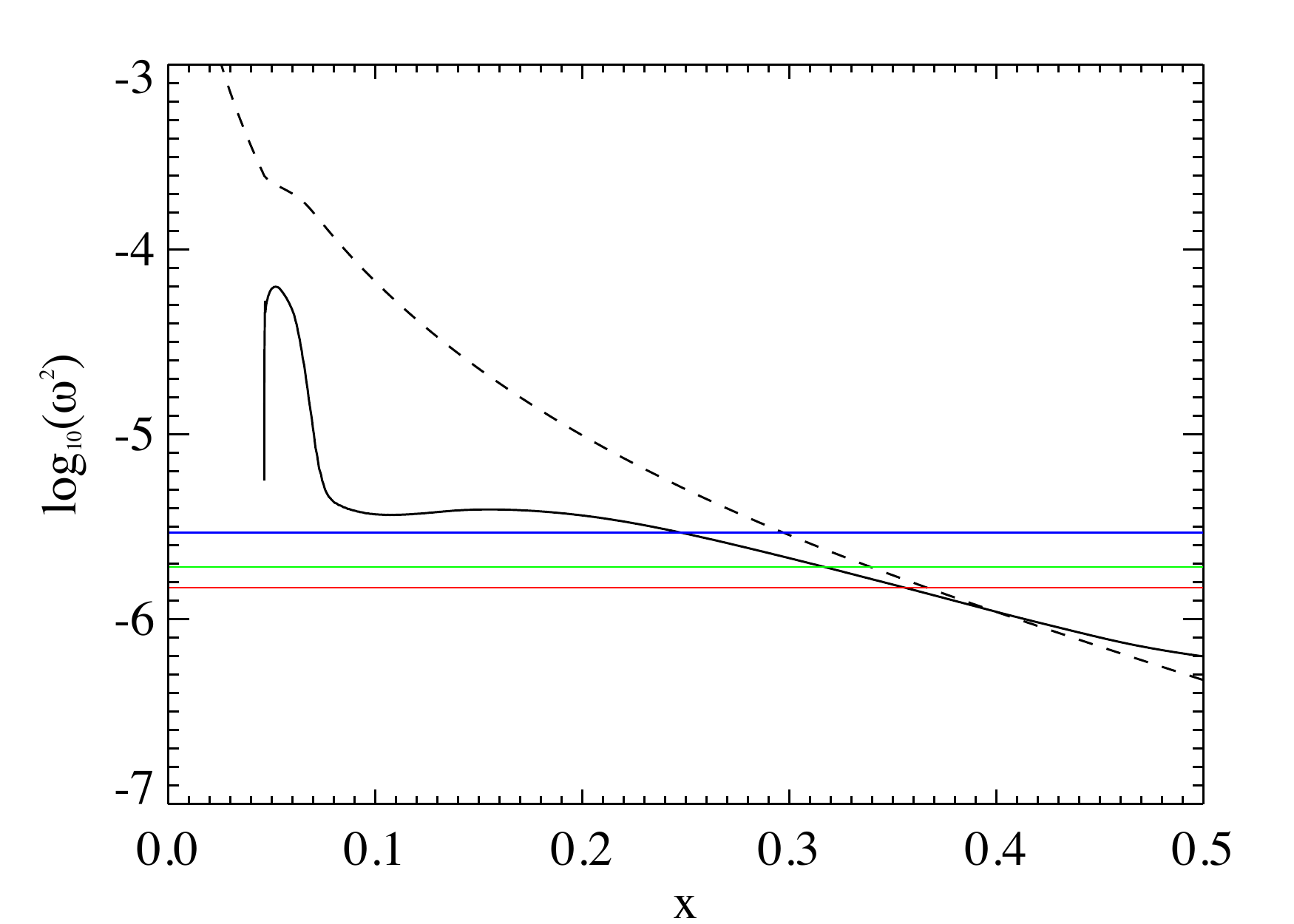}
\caption{\footnotesize Propagation diagram for the three quintuplets identified as mixed modes. The frequencies of the blue, green, and red lines are 23.565 $\rm{d}^{-1}$, 19.006 $\rm{d}^{-1}$, and 16.742 $\rm{d}^{-1}$, respectively. The solid black line represents the Brunt-$\rm{V}\ddot{a}is\ddot{a}l\ddot{a}$ frequency $N^2$ and the dashed line represents the Lamb frequency  $S_{2}^2$, which is computed setting $l=2$. We see that the evanescent region (defined as a region where $N^2 < \omega^2$ and $S_{2}^2 > \omega^2$, or, $N^2 > \omega^2$ and $S_{2}^2 < \omega^2$) is narrow enough for the three modes to be mixed modes.}
\label{propagation_diagram_l2_expanded}
\end{center} 
\end{figure}
Finally, we would like to discuss the model dependence of the inferred fast-core rotation. As we have shown in Section \ref{5-1}, our inference of the fast-core rotation is strongly dependent on sensitivity of the mixed-mode splitting kernel with $(n,l)=(2,2)$. However, we have found that the fast-core rotation is inferred even when we exclude the mode $(2,2)$ in the three-zone modeling inversion, suggesting that the other mixed modes with $l=2$ also sense inside the convective core. Thus, the point is whether the modes identified as mixed modes are really mixed modes or not. When we look at the propagation diagram (Figure \ref{propagation_diagram_l2_expanded}), we see that the Brunt-$\rm{V}\ddot{a}is\ddot{a}l\ddot{a}$ frequency should be smaller in order for the three $l=2$ modes to be identified as not mixed modes because we need a broad evanescent region between the g-mode cavity and the p-mode cavity. But the Brunt-$\rm{V}\ddot{a}is\ddot{a}l\ddot{a}$ frequency is well constrained by the observed value of the mean g-mode period spacing \citep{Kurtz2014} and there is little room for modifying it. Thus it is likely that the modes in the frequency range are mixed modes, leading to the robustness of the inferred fast-core rotation.

\subsection{A broad picture of the internal rotation of KIC11145123}
\label{5-3}
We would like to provide some physical considerations concerning the inferred internal rotation. One is the internal rotation of the radiative region of the star where the envelope rotates slightly faster than the deep radiative region does. This trend was first found by \citet{Kurtz2014} and we have confirmed it in a two-dimensional way (see Section \ref{5-1}). It is apparently peculiar that the envelope rotates faster than the deep radiative region does since the angular momentum redistribution by viscosity alone cannot produce such an internal rotation profile. \citet{Rogers2015} has carried out two-dimensional numerical simulations of angular momentum transfer by internal gravity waves, and discussed that it is possible that such a mechanism renders the outer envelope rotate faster than the inner region, which is the case for KIC11145123. \citet{Benomar2015}, however, did not find any evidence for such a reversed trend of stellar rotation via the measurements of radial differential rotation of solar-like stars. Then, another possible hypothesis is that the star has experienced some interactions such as mass accretion or stellar collision with other stars during the evolution and it has obtained angular momentum from the outside, leading to the spin-up of the outer envelope. Actually, \citet{Takada-Hidai2017} have shown that the star is spectroscopically a blue straggler, which is thought to be born via interactions with other stars; their results support the above explanation. If this is really the case, our results could be a constraint on the theory of stellar interactions in a binary system. 

The other is a strong shear of the angular velocity at the boundary between the convective core and the radiative region above. It is usually expected that such a velocity shear causes instabilities between the two boundaries and it leads to an extra mixing there. Interestingly enough, fine-tuning of the model of the star suggests that the atomic diffusion processes in the deep radiative region should be much weaker than usually thought (Hatta et al., in prep.) in typical stellar evolution codes; the extra mixing caused by the shear could counteract atomic diffusion processes. This mechanism is rather similar to that occurring at the solar tachocline. 

\section{Summary}
\label{6}
A detailed asteroseismic perturbative approach for inferring the internal rotation of a possible blue straggler KIC11145123 has provided us with two fascinating aspects of the interior rotation of the star. 

One is the radiative region ($0.045<x<1.0$) which rotates slowly ($P\sim 104.6 \ \rm{d}$) and almost rigidly throughout the region. However, there definitely exists differential rotation in the star. For the outer radiative region ($x\sim0.95$), the latitudinally differential rotation has been detected with a marginal significance based on the model-independent consideration, which, in addition, favors that the high-latitude region rotates faster than the low-latitude region does. This is the first time such latitudinally differential rotation of an A-type star has been inferred based on the asteroseismic analysis. It is also found that the deep radiative region ($x\sim0.05$) rotates slightly slower ($P\sim 105.3 \ \rm{d}$) than the outer radiative region ($x\sim0.95$) does ($P \sim 104.1 \ \rm{d}$). This is essentially the same trend that \citet{Kurtz2014} previously found. One possible explanation for such an unusual spin-up of the outer region is that the star has obtained extra angular momentum from the outside during the evolution by some interactions with other stars. 

On the contrary to the almost uniform rotation in the radiative region, the convective core rotates about 6 times faster than the other parts of the star. This is also the first case the internal rotation of the convective core has been inferred and a strong velocity shear between the convective core and the radiative region above has been suggested. The shear of the angular velocity might be a cause of an extra mixing process which is suggested by the fine-tuning of the model of the star.

Our results are definitely giving a new insight into the understanding of the internal rotation of main-sequence stars. However, these results have been obtained based on the assumption that Kurtz et al.'s model, which cannot reproduce the observed p-mode frequencies, is the best model of the star. We have confirmed the same trend of the internal rotation for several other best models suggested by other authors \citep[e.g.][]{Takada-Hidai2017}, but still, we have not been able to completely eliminate the dependence of our inferences on the select of equilibrium models of the star. To find the better model of the star, which was once thought to be a main-sequence A-type star but recently has been proposed to be a blue straggler based on spectroscopy, non-standard modeling of the star is necessary, and that will eventually lead us for comprehending the more accurate description of the internal rotation of the star as well as of the physical properties of it. This is our next goal to be achieved.

%
%
%
\ 

The authors would like to thank NASA and Kepler team for their invaluable data. We thank H. Saio for his helpful and insightful advices. H. Shibahashi is also thanked for his constructive comments.

\end{document}